\documentclass[reprint,superscriptaddress,amsmath,amssymb,nolongbibliography,aip,jcp,floatfix]{revtex4-2}

\usepackage[acronym,toc,nonumberlist,nopostdot,nogroupskip]{glossaries}
\usepackage{hyperref}
\hypersetup{
    breaklinks=true,
    colorlinks=true,
    linkcolor=blue,
    filecolor=red,
    urlcolor=blue,
    pdftitle={ScaLAPACK},
    pdfpagemode=FullScreen,
}
\usepackage{xspace}
\usepackage[table]{xcolor}
\usepackage{amsmath,amssymb,graphicx,bm}
\usepackage{braket}
\usepackage[inline]{enumitem}
\usepackage{booktabs}
\usepackage{array}
\usepackage[version=4]{mhchem}
\usepackage[super]{nth}

\newacronym{BZ}{BZ}{first Brillouin zone}
\newacronym{DFT}{DFT}{density functional theory}
\newacronym{DFPT}{DFPT}{density functional perturbation theory}
\newacronym{PES}{PES}{potential energy surface}
\newacronym{BEC}{BEC}{Born effective charge}
\newacronym{EDA}{EDA}{electric dipole approximation}
\newacronym{ML}{ML}{machine-learning}
\newacronym{ENN}{ENN}{equivariant neural network}
\newacronym{FPS}{FPS}{farthest point sampling}
\newacronym{MLIP}{MLIP}{machine-learning interatomic potential}
\newacronym{MD}{MD}{molecular dynamics}
\newacronym{PIMD}{PIMD}{path integral molecular dynamics}
\newacronym{MSD}{MSD}{mean square displacement}
\newacronym{NAO}{NAO}{numeric atomic orbital}
\newacronym{BLACS}{BLACS}{Basic Linear Algebra Communication Subprograms}
\newacronym{ELSI}{ELSI}{Electronic Structure Infrastructure}
\newacronym{mpi}{MPI}{message-passing interface}
\newacronym{MTP}{MTP}{modern theory of polarization}
\newacronym{soc}{SOC}{spin-orbit coupling}
\newacronym{xc}{xc}{exchange and correlation}

\newglossaryentry{pbc}{
    name=periodic boundary conditions,
    description={periodic boundary conditions}
}

\newglossaryentry{BvK}{
    name=Born--von Karman,
    description={Born--von Karman}
}

\newglossaryentry{KS}{
    name=Kohn--Sham,
    description={Kohn--Sham}
}

\newglossaryentry{oxn}{
    name=oxidation number,
    description={oxidation number}
    plural={oxidation number}
}

\newglossaryentry{puc}{
    name=primitive unit cell,
    description={primitive unit cell}
}

\newglossaryentry{ae}{
    name=atomic environment,
    description={atomic environment}
    plural={atomic environment}
}

\newglossaryentry{BO}{
    name=Born-Oppenheimer,
    description={Born-Oppenheimer}
}

\newglossaryentry{ham}{
    name=Hamiltonian,
    description={Hamiltonian}
}

\newcolumntype{R}[1]{>{\raggedleft\arraybackslash}p{#1}}
\newcolumntype{L}[1]{>{\raggedright\arraybackslash}p{#1}}
\newcolumntype{C}[1]{>{\centering\arraybackslash}p{#1}}

\newlist{mylist}{enumerate*}{1}
\setlist[mylist]{label=(\roman*)}
\setlist[itemize]{itemsep=0pt}

\newcommand{\supercell}[1]{\ensuremath{#1\times#1\times#1}}

\newcommand{\nelec}{\ensuremath{N_{\rm e}}\xspace}
\newcommand{\nnuc}{\ensuremath{N_{\rm n}}\xspace}
\newcommand{\wavefunc}{\ensuremath{\Psi_n^{\bm{k}}}\xspace}
\newcommand{\perfunc}{\ensuremath{u_n^{\bm{k}}}\xspace}
\newcommand{\der}[2]{\ensuremath{ \nabla_{#2} #1 }\xspace}

\newcommand{\aims}{\texttt{FHI-aims}\xspace}
\newcommand{\ie}{i.e.,~}
\newcommand{\eg}{e.g.,~}
\newcommand{\plane}{\ensuremath{\mathcal{T}_{\perp,\alpha}}\xspace}
\newcommand{\myloop}{\ensuremath{\mathcal{L}_{\alpha}(\bm{k}_{\perp,\alpha})}\xspace}

\newcommand{\dipmat}{\ensuremath{\mathcal{D}_{\nu\mu}^{\bm{k},\alpha}}\xspace}
\newcommand{\dipmatR}{\ensuremath{\mathcal{D}_{\nu\mu}^{\bm{L},\alpha}}\xspace}

\newcommand{\dipoleterm}{\ensuremath{\mathfrak{D}^\alpha_n(\bm{k})}\xspace}
\newcommand{\berryterm}{\ensuremath{\mathfrak{B}^\alpha_n(\bm{k})}\xspace}

\newcommand{\myline}[1]{\arrayrulecolor{black!80}
\cmidrule(l{0pt}r{0pt}){#1}
\arrayrulecolor{black}}

\makeatletter
\DeclareRobustCommand{\fig}[1]{%
  \@ifundefined{fig@#1}{%
    \expandafter\gdef\csname fig@#1\endcsname{}%
    Figure~\ref{#1}%
  }{Figure~\ref{#1}}%
}
\makeatother

\makeatletter
\DeclareRobustCommand{\tab}[1]{%
  \@ifundefined{tab@#1}{%
    \expandafter\gdef\csname tab@#1\endcsname{}%
    Table~\ref{#1}%
  }{Table~\ref{#1}}%
}
\makeatother

\DeclareRobustCommand{\kpoint}{\ensuremath{\bm{k}}\nobreakdash-point\xspace}
\DeclareRobustCommand{\kpoints}{\ensuremath{\bm{k}}\nobreakdash-points\xspace}
\DeclareRobustCommand{\lapack}{\texorpdfstring{\texttt{LAPACK}}{LAPACK}\xspace}
\DeclareRobustCommand{\scalapack}{\texorpdfstring{\texttt{ScaLAPACK}}{ScaLAPACK}\xspace}
\DeclareRobustCommand{\MTPRefs}{\cite{SmithVanderbilt1993,Resta1994,Resta2007,Resta2022SinglePoint}}
\DeclareRobustCommand{\bto}{\ce{BaTiO3}\xspace}

\DeclareRobustCommand{\scf}{\texttt{scf}\xspace}
\DeclareRobustCommand{\nscf}{\texttt{nscf}\xspace}



\setcitestyle{doi=false}
\setcitestyle{url=false}
\setcitestyle{eprint=false}

\DeclareRobustCommand{\forcecaptionleft}{%
    \ifinner
        \hspace{\linewidth}%
    \fi
}

\begin{document}

    \title{Scalable Berry-Phase Polarization Calculations with Numeric Atomic Orbitals}

        \let\oldaffiliation\affiliation
        \renewcommand{\affiliation}[1]{\oldaffiliation{\mbox{#1}}}

        \author{Elia Stocco}
        \affiliation{MPI for the Structure and Dynamics of Matter, Luruper Chaussee 149, 22761 Hamburg, Germany}

        \author{James A. Green}
        \affiliation{Molecular Simulations from First Principles e.V., Akazienstr. 3A, 10823, Berlin, Germany}

        \author{Mariana Rossi}
        \email[Co-corresponding Author: ]{mariana.rossi@mpsd.mpg.de}
        \affiliation{MPI for the Structure and Dynamics of Matter, Luruper Chaussee 149, 22761 Hamburg, Germany}
        \affiliation{Yusuf Hamied Department of Chemistry, University of Cambridge, Lensfield Road, Cambridge CB2 1EW, UK}

        \author{Christian Carbogno}
        \email[Corresponding Author: ]{carbogno@fhi-berlin.mpg.de}
        \affiliation{Theory Department, Fritz Haber Institute of the Max Planck Society, Faradayweg 4-6, 14195 Berlin, Germany}

    \begin{abstract}
        \noindent
        Obtaining the polarization of crystalline materials from first-principles electronic structure calculations requires evaluating the geometric Berry phase, as formalized by the modern theory of polarization.
        Existing implementations of this formalism are predominantly based on plane-wave architectures, which can limit scalability both to large core counts and to large systems featuring many atoms.
        Here, we present a density-functional theory implementation of the polarization using the Berry-phase formalism in the all-electron \aims code.
        We discuss formal and algorithmic adaptations that enable highly parallel calculations for large systems within this implementation based on numeric atom-centered orbitals.
        In particular, this is enabled by expressing the Berry connection through a translationally invariant real-space dipole matrix, evaluated with the same integration routines used for the Hamiltonian and overlap matrices, and a finite-difference term that requires only the Kohn--Sham coefficients at neighboring $\bm{k}$-points and the overlap matrix.
        This avoids $\bm{k}$-dependent phase-operator integrals and is not restricted to $\Gamma$-point sampling.
        Distributing the matrices of each $\bm{k}$-point over groups of cores with \scalapack removes the memory and parallelization bottlenecks of a $\bm{k}$-point-only distribution, which is decisive for large systems.
        We demonstrate the validity of the implementation for several test systems and show that it achieves near-ideal parallel scalability and excellent scaling for systems containing more than 1,000 atoms.
    \end{abstract}

    \maketitle

    \section{Introduction} \label{sec:intro}

    The polarization and its response to changes in system parameters, such as nuclear coordinates and strains, are fundamental material properties of great importance in fields such as electrochemistry~\cite{Sprik2020,Wippermann2021}, piezoelectrics~\cite{Cohen2008}, ferroelectrics~\cite{RobertoCarPbTiO3,StoccoRossi2025ml,RobertoCarFES,ferroelectricreview}, and multiferroics~\cite{Spaldin2026,fechnerdynpol}.
    However, the computational assessment of such properties using electronic structure codes has not yet reached the same level of maturity, performance, and scalability as calculations of total energies and atomic forces.
    For instance, \gls{DFT}~\cite{dft1964} calculations based on the \gls{KS} formalism~\cite{KS1965} are nowadays capable of addressing systems containing thousands of atoms~\cite{roadmapDFT}, even when hybrid functionals are used~\cite{FHIaims2024}, whereas polarization calculations are often limited to systems with hundred atoms or even less.

    One fundamental reason for this situation is that evaluating the polarization of periodic systems within the \gls{MTP} is computationally more demanding and conceptually more involved than standard total-energy and force calculations~\MTPRefs.
    Moreover, this formulation poses additional challenges for efficient distributed-memory parallelization, and existing implementations are often limited in their scalability and therefore in the system sizes they can address.

    In this work, we present an implementation for calculating the polarization in \aims~\cite{FHIaims2009,FHIaims2024,FHIaimsroadmap2025} that overcomes these limitations.
    In particular, we employ a \scalapack~\cite{scalapack} scheme to distribute both the computational workload and memory requirements, thereby enabling calculations for systems containing thousands of atoms.
    This capability facilitates the construction of large, accurate datasets of polarization-related properties for training machine-learning models for the electronic structure~\cite{MLroadmap2022}.

    This work is organized as follows.
    Section~\ref{sec:theory} briefly presents the \gls{MTP} in periodic systems, with a focus on its implications for localized atomic-orbital implementations.
    Section~\ref{sec:practical} presents the computational and infrastructural details of the current implementation.
    Section~\ref{sec:results} presents the results of the validation tests and demonstrates the scalability of the code.
    Section~\ref{sec:conclusion} provides concluding remarks.

    \section{Theoretical Background}\label{sec:theory}

    In this section, we present the mathematical expressions implemented in the code to calculate the polarization of periodic systems.
    First, Sec.~\ref{subsec:MTP} introduces the central concepts of the \gls{MTP} and explains the fundamental relation between the polarization of a system and its electronic Berry connection.
    Next, Sec.~\ref{subsec:KS-NAO} addresses the practical aspects of the \gls{MTP} when the electronic structure is represented in a localized, potentially non-orthogonal basis set.
    Finally, Sec.~\ref{sec:numerical} derives the numerical formulation on finite \kpoint meshes and discusses its gauge invariance.
    \subsection{Modern Theory of Polarization}\label{subsec:MTP}
    In this work, we consider a periodic system described by a unit cell with lattice vectors $(\bm{a}_1,\bm{a}_2,\bm{a}_3)$.
    Each unit cell contains \nnuc nuclei with positions $\bm{R}_I$ and charges~$Z_I$ as well as \nelec electrons.
    For simplicity, spin degrees of freedom are omitted in this notation at first; they will then be reintroduced in Sec.~\ref{SecSpin}.
    In such a periodic system, the electronic wavefunction can be expressed in the \textit{crystal momentum representation}~\cite{CMR,blount} using Bloch functions~\cite{bloch1929}.
    For instance, the individual \gls{KS} states~$\wavefunc(\bm{r})$ used in \gls{KS} \gls{DFT}~\cite{dft1964,KS1965} can be written as
    \begin{align}
        \wavefunc(\bm{r}) = e^{i\bm{k}\cdot\bm{r}} \perfunc(\bm{r}).
        \label{eq:bloch}
    \end{align}
    Here, $\bm{k}$ is a reciprocal-space vector, $n$ is the band index, and $\perfunc$ is a lattice-periodic function that satisfies $\perfunc(\bm{r}+\bm{L}) = \perfunc(\bm{r})$ for any combination of lattice vectors~$\bm{L}$.
    The wavefunctions $\wavefunc(\bm{r})$ are normalized over the whole volume of the \gls{BvK} supercell.
    Along the same lines, one typically imposes the \textit{periodic gauge} condition, \ie
    \begin{align}
        \Psi_n^{\bm{k}+\bm{G}}(\bm{r}) = \wavefunc(\bm{r})
        \label{eq:periodic-gauge}
    \end{align}
    for any combination of reciprocal-lattice vectors $\bm{G}$, which in turn implies $u_n^{\bm{k}+\bm{G}}(\bm{r}) = e^{-i\bm{G}\cdot\bm{r}} \perfunc(\bm{r})$.

\begin{figure}[!t]
    \centering
    \includegraphics[width=0.8\linewidth]{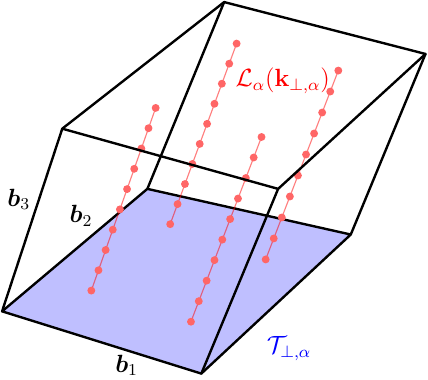}
    \caption{Sketch illustrating the $\bm{k}$-point meshes used to perform integrals over the reciprocal space.
    The integrals along $\bm{b}_{3}$ are performed as closed one-dimensional line integrals, referred to as ``strings'', which yield gauge-invariant results with the formalism proposed in the \gls{MTP}.
    Accordingly, the integral over the manifold~$\mathcal{T}_{\perp,3}$ described by the remaining dimensions~$\bm{b}_{1}$ and~$\bm{b}_{2}$, here marked by the blue plane at $k_{3}=0$, can then be performed by standard numerical techniques.}
    \label{fig:strings}
\end{figure}

    According to the \gls{MTP}~\MTPRefs, the electric polarization $\bm{P} = \bm{P}_n + \bm{P}_e$ of such a system comprises a nuclear contribution $\bm{P}_n$ and an electronic contribution $\bm{P}_e$, which are defined as:
    \begin{align}
        \bm{P}_n
            = & \,
                \frac{e}{\Omega} \sum_I^{\nnuc} Z_I \bm{R}_{I} \\
        \bm{P}_e
            = & \,
               - \frac{e}{\left(2\pi\right)^3}
                \sum_n^{\rm occ} \oint_{\rm BZ} \, d^3\bm{k} \,
                \,
                \bm{\mathcal{A}}_{n}\left(\bm{k}\right).  \label{eq:polarization-electronic}
    \end{align}
    Here, $e$ is the elementary charge, $\Omega$ is the volume of the unit cell, the sum runs over all occupied states, and the reciprocal-space integral is performed over the \gls{BZ}.
    Furthermore, $\bm{\mathcal{A}}_{n}\left(\bm{k}\right)$ is the Berry connection of the $n^{\rm th}$ band, defined as
    \begin{align}
        \bm{\mathcal{A}}_{n}\left(\bm{k}\right)
            =  \,
                i \braket{\perfunc | \der{\perfunc}{\bm{k}} } \label{eq:berry-connection}
            =  \,
                \sum_{\alpha=1,2,3}
                \frac{\bm{a}_\alpha}{2 \pi} \,
                \underbrace{
                    i \braket{\perfunc | \frac{ \partial \perfunc}{\partial k_{\alpha}} }
                }_{\mathcal{A}_{n}^{\alpha}\left(\bm{k}\right)}.
    \end{align}
    In the last step, we explicitly separated the Berry connection into components~$\mathcal{A}_{n}^{\alpha}\left(\bm{k}\right)$ along the reciprocal-lattice vectors~$\bm{b}_{\alpha}$ using $\bm{k}=\sum_{\alpha=1,2,3} k_\alpha \bm{b}_{\alpha}$, with $k_\alpha\in[0,1)$ (see the supplementary material, Sec.~S1).
    This separation allows us to recast the polarization and the reciprocal-space integral in Eq.~\eqref{eq:polarization-electronic} for each direction~$\alpha$ in terms of the component $P_e^\alpha$, defined by $\bm{P}_e=\sum_{\alpha=1,2,3}P_e^\alpha\bm{a}_\alpha/|\bm{a}_\alpha|$, as
    \begin{align}
        P^{\alpha}_e
            = & \,
                - \frac{e}{\left(2\pi\right)^3}
                \sum_n^{\rm occ}
                \phi_n^{\alpha}  \label{eq:berry-polarization}\\
        \phi_n^{\alpha}
            = & \,
                \oint_{\plane}
                d^2 \bm{k}_{\perp,\alpha} \,
                \phi_n(\bm{k}_{\perp,\alpha})  \label{eq:tot-berry-phase} \\
            \phi_n(\bm{k}_{\perp,\alpha})
            = & \,
                \oint_{\myloop}
                dk_{\alpha}\,
                \mathcal{A}_{n}^{\alpha}\left(\bm{k}\right).  \label{eq:berry-phase}
    \end{align}
    We recognize that the last equation has the mathematical form of a Berry phase~\cite{berry1984}.
    In these equations, we have expressed the component $P^{\alpha}_e$ of the electronic polarization along the lattice vector $\bm{a}_\alpha$ by writing $\bm{k} = \bm{k}_{\perp,\alpha} + k_{\alpha}\bm{b}_{\alpha}$ in terms of the coordinates $k_{\alpha}$ and $\bm{k}_{\perp,\alpha}$, which are parallel and perpendicular, respectively, to the reciprocal-space vector~$\bm{b}_{\alpha}$.
    The reciprocal-space integral in Eq.~\eqref{eq:polarization-electronic} has been decomposed into a family of closed one-dimensional line integrals along the strings \myloop parallel to $\bm{b}_{\alpha}$ and a two-dimensional surface integral over the transverse manifold $\plane$~\MTPRefs.
    This separation into surface and line integrals, also represented in \fig{fig:strings}, is advantageous because it simplifies the numerical treatment of gauge invariance for the following reasons.
    The wavefunctions $\wavefunc$ and $\perfunc$ are gauge dependent, since they can be multiplied by an arbitrary phase factor $e^{i\eta_n(\bm{k})}$ without changing their physical content; consequently, the Berry connection $\bm{\mathcal{A}}_{n}\left(\bm{k}\right)$ is also gauge dependent.
    However, the periodic-gauge condition in Eq.~\eqref{eq:periodic-gauge} ensures that the line integral in Eq.~\eqref{eq:berry-phase} leads to a gauge-invariant Berry phase~\cite{berry1984}.
    The decomposition in Eqs.~\eqref{eq:tot-berry-phase} and~\eqref{eq:berry-phase} therefore expresses the polarization as a surface integral over gauge-invariant Berry phases,
    which can be evaluated using standard integration techniques.
    Therefore, the mentioned decomposition allows preserving gauge invariance by enforcing the periodic-gauge condition only along each one-dimensional string rather than throughout the full three-dimensional \gls{BZ}.
    Section~\ref{sec:numerical} describes how this condition is enforced on a finite \kpoint grid.

    Despite their gauge invariance, the topology of the integration domain implies that the Berry phases $\phi_n(\bm{k}_{\perp,\alpha})$ in Eq.~\eqref{eq:berry-phase} are only defined modulo $2\pi$~\MTPRefs.
    This $2\pi$ ambiguity carries over to the polarization: each component $P^\alpha_e$ is multivalued, \ie defined modulo an integer multiple of the polarization quantum $Q_\alpha=e|\bm{a}_\alpha|/\Omega$~\MTPRefs.
    The resulting set of values, separated by integer multiples of $Q_\alpha$, constitutes the different \textit{branches} of the polarization.
    When comparing the polarization obtained for different structures or from distinct calculations, one must ensure that the corresponding values lie on the same branch.
    In practice, this is achieved by \textit{branch unfolding}~\cite{Spaldin2012}, which is a procedure that can be summarized as follows: (i) a continuous path connecting the two structures is defined; (ii) the polarization is evaluated at sufficiently closely spaced points along this path to unambiguously identify the branch of each successive value; and (iii) integer multiples of the polarization quantum are added or subtracted as needed to obtain a continuous evolution of the polarization along the path.

    A quantity closely related to the polarization is the \gls{BEC} $Z^{*I}_{\alpha\beta}$ of the $I^{\rm th}$ nucleus.
    Although the polarization itself is multivalued, its derivatives are uniquely defined and do not suffer from the indeterminacy associated with the polarization quantum.
    The \glspl{BEC} are defined as the derivative of the polarization component $P^{\alpha}$ with respect to the nuclear position Cartesian components $R_{I}^{\beta}$~\cite{Resta2007,Spaldin2012,vanderbilt2018}:
    \begin{align}
        Z^{*I}_{\alpha\beta}
            = & \,
                \Omega \, \frac{ \partial P^{\alpha}}{\partial R_{I}^{\beta}} \label{eq:bec} .
    \end{align}
    \Glspl{BEC} are fundamental quantities in the study of lattice dynamics and dielectric properties of materials.
    As they describe the response of the polarization to atomic displacements, they determine the infrared activity of phonon modes~\cite{Vibes} and the splitting between longitudinal and transverse optical modes in polar crystals~\cite{dfptbec}.
    Moreover, they are essential for describing nuclear motion in the presence of external electric fields~\cite{StoccoRossi2025ml,cassone,futera,Vanderbilt2002,Pasquarello2002}.
    While formulations based on \gls{DFPT} allow \glspl{BEC} to be computed analytically~\cite{dfptbec}, this particular capability is not yet available in the \gls{DFPT} implementation~\cite{Shang.2017,Shang.2018ucl} of \aims.
    \glspl{BEC} can nevertheless be obtained from finite differences of the polarization according to Eq.~\eqref{eq:bec}, as done in Sec.~\ref{sec:results} of this work for validation purposes.

    \subsection{Formulation for Localized Atomic Orbitals}\label{subsec:KS-NAO}

    In this section, we derive practical expressions for evaluating the polarization within the \gls{MTP} using localized, potentially non-orthogonal orbitals~\cite{FHIaimsroadmap2025}.
    The implementation considered here uses \glspl{NAO} in the all-electron electronic structure code~\aims~\cite{FHIaims2009,FHIaims2024,FHIaimsroadmap2025}.
    Notably, most implementations of the \gls{MTP} are based on plane-wave codes, including \texttt{Quantum ESPRESSO}~\cite{qe2020}, \texttt{ABINIT}~\cite{abinit2020}, and \texttt{VASP}~\cite{vasp1994,vasp1996,vasp1996bis}.
    Existing implementations employing localized basis sets use approaches different from the one used in \aims.

    For instance, \texttt{CRYSTAL} evaluates the discretized Berry phase from determinants of overlap matrices between the occupied orbitals at neighboring \kpoints~\cite{crystalberry1,crystalberry2,crystal2023}.
    Constructing these cross-$\bm{k}$ overlaps requires matrix elements of the phase operator $e^{-i\Delta\bm{k}\cdot\bm{r}}$ between basis functions, which can be evaluated analytically in its localized Gaussian basis.
    \texttt{SIESTA} follows the same approach but, because it uses numerical pseudo-atomic rather than Gaussian orbitals, approximates these integrals by expanding the phase operator to first order in $\Delta\bm{k}$~\cite{siesta2002,siesta2020}.
    Conversely, \texttt{CP2K}~\cite{CP2K} implements the \gls{MTP} in the $\Gamma$-only case by evaluating the polarization as $P^\alpha = {\rm Im} \ln \det \mathcal{Q}^\alpha$ with $\mathcal{Q}^\alpha_{mn} = \braket{\Psi_m|\exp(i\bm{b}_\alpha\bm{r})|\Psi_n}$, as originally proposed by Resta in Ref.~\cite{Resta1998}.
    Although this approach can be generalized to arbitrary \kpoint grids~\cite{Resta1998}, doing so would require the explicit evaluation of numerous $\bm{k}$-dependent real-space integrals and would be numerically disadvantageous with a localized basis set.

    Conversely, the formulation used in this work treats general \kpoint grids directly in a localized \gls{NAO} basis by separating the Berry connection into contributions from translationally invariant dipole matrix elements and from $\bm{k}$-derivatives of the \gls{KS} coefficients, thereby avoiding both explicit $\Delta\bm{k}$-dependent phase-operator integrals and the $\Gamma$-point restriction, as derived in detail below.
    We thus overcome the limitations described above and lay the foundations for a general, scalable, and parallel implementation of the Berry connection for localized basis sets.

    \begin{figure}[!t]
    \centering
    \includegraphics[width=\linewidth]{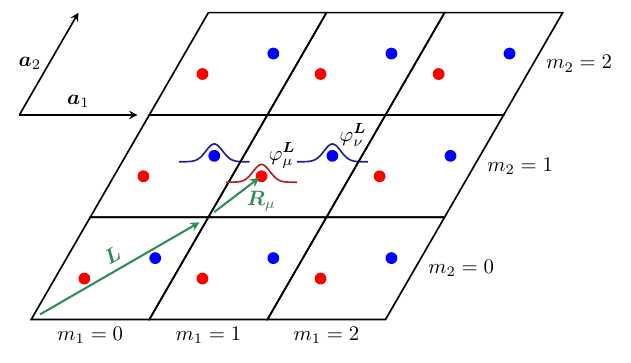}
    \caption{Sketch of localized orbitals in a 2D $3\times3$ \gls{BvK} supercell and of the associated \glspl{NAO} $\varphi_\mu^{\bm{L}}$.
    The red and blue dots represent the periodic images of two atoms, while $\bm{L}$ and $\bm{R}_\mu$ show how the center $\bm{L}+\bm{R}_\mu$ of $\varphi_\mu^{\bm{L}}$ is constructed.}
    \label{fig:BvK}
\end{figure}

    To formulate our implementation, we first describe the representation of the electronic structure in \aims.
    The code solves the \gls{KS} equations by expanding the single-particle wavefunctions in a basis of \glspl{NAO}, \ie localized orbitals centered on the nuclei with numerically tabulated radial parts.
    In periodic systems, the basis function $\varphi_{\mu}^{\bm{L}}$ in the \gls{BvK} supercell is defined as
    \begin{align}
        \varphi_{\mu}^{\bm{L}}(\bm{r}) =  & \,
        \varphi_{\mu}(\bm{r}-\bm{R}_{\mu}-\bm{L}) .\label{eq:basis-r-space}
    \end{align}
    Here, $\bm{R}_{\mu}$ is the position of the atom $\mu$ on which the function is centered in the reference unit cell, and $\bm{L}$ is a lattice vector within the \gls{BvK} supercell.
    Thus, $\varphi_{\mu}^{\bm{L}}$ is centered at $\bm{R}_{\mu}+\bm{L}$, as depicted in \fig{fig:BvK}.
    To exploit the translational symmetries of periodic systems, these \glspl{NAO} can be used to construct Bloch-like reciprocal-space orbitals $\chi_{\mu}^{\bm{k}}$ as
    \begin{align}
        \chi_{\mu}^{\bm{k}}(\bm{r}) = & \,
        \frac{1}{\sqrt{N}}
        \sum_{\bm{L}} \,
        e^{i \bm{k}\cdot\bm{L}} \,
        \varphi_{\mu}^{\bm{L}}(\bm{r}).  \label{eq:basis-k-space}%
    \end{align}
    Here, $N$ is the number of unit cells in the \gls{BvK} supercell.
    This definition of $\chi_{\mu}^{\bm{k}}$ is usually referred to in the literature as the \textit{lattice gauge}, since only the lattice vectors, but not the actual positions in the unit cell, enter the Bloch exponential~\cite{atomicgaugescipost}.
    An alternative convention, known as the \textit{atomic gauge}, incorporates the positions $\bm{R}_\mu$ into the phase factors and is discussed in Appendix~\ref{sec:berry-atomic}.
    Both gauges are implemented in \aims for the calculation of the polarization and yield the same physical observables.
    For clarity, we focus on the lattice gauge throughout the main derivation.

    The single-particle \gls{KS} wavefunctions $\wavefunc$ are expanded in this basis set using the coefficients $C_{\mu n}^{\bm{k}}$:
    \begin{align}
        \wavefunc(\bm{r}) =
        \sum_{\mu}^{N_{\rm basis}} \,
        C_{\mu n}^{\bm{k}} \,
        \chi_{\mu}^{\bm{k}}(\bm{r}) .  \label{eq:expansion}
    \end{align}

        Since the underlying \glspl{NAO} are non-orthogonal, the resulting algebraic \gls{KS} equations become a generalized eigenvalue problem~\cite{FHIaims2009}:
        \begin{align}
            \sum_{\mu}^{N_{\rm basis}} \,
            \mathcal{H}^{\bm{k}}_{\nu\mu}
            \,
            C_{\mu n}^{\bm{k}}
            = & \,
            \varepsilon_{n}^{\bm{k}}  \,
            \sum_{\mu}^{N_{\rm basis}} \,
            \mathcal{S}_{\nu\mu}^{\bm{k}}
            \,
            C_{\mu n}^{\bm{k}}
        \end{align}
        where $\varepsilon_{n}^{\bm{k}}$ are the \gls{KS} eigenvalues, and $\mathcal{H}^{\bm{k}}_{\nu\mu}$ and $\mathcal{S}_{\nu\mu}^{\bm{k}}$ are the elements of the Hamiltonian and overlap matrices in the $\chi_{\mu}^{\bm{k}}$ representation, respectively (see the supplementary material, Sec.~S2 for mathematical details).

    To derive an expression for the Berry connection in terms of localized orbitals, we insert the lattice-periodic wavefunction $\perfunc(\bm{r})=e^{-i\bm{k}\cdot\bm{r}}\wavefunc(\bm{r})$ into Eq.~\eqref{eq:berry-connection}, which gives
    \begin{align}
        \mathcal{A}^{\alpha}_{n}\left(\bm{k}\right)
            =
            \braket{ \Psi_{n\bm{k}} \,|\,  r^\alpha  \,|\, \Psi_{n\bm{k}} }
            + i
            \braket{ \Psi_{n\bm{k}}  \,|\,  \frac{\partial \Psi_{n\bm{k}}  }{\partial k_\alpha } }
            \label{eq:berry-connection-psi}.
    \end{align}

    Here, $r^{\alpha}\equiv\partial(\bm{k}\cdot\bm{r})/\partial k_{\alpha}=\bm{b}_{\alpha}\cdot\bm{r}$ denotes the component associated with $\bm{a}_{\alpha}$.
    The scalar product with the reciprocal lattice vector $\bm{b}_{\alpha}$, rather than with $\bm{a}_{\alpha}$, is required because the direct lattice vectors are generally non-orthogonal (see the supplementary material, Sec.~S1).
    Accordingly, we use the analogous notation $\mathcal{R}_{\mu}^{\alpha} \equiv \bm{b}_{\alpha}\cdot\bm{R}_{\mu}$ and $L^{\alpha} \equiv \bm{b}_{\alpha}\cdot\bm{L}$ in the following.
    The first term of Eq.~\eqref{eq:berry-connection-psi}, \ie the expectation value of the position operator in the crystal-momentum representation~\cite{CMR,blount}, can be expanded in the localized basis functions using Eq.~\eqref{eq:expansion} leading to
    \begin{align}
        \braket{ \Psi_{n\bm{k}} \,|\,  r^{\alpha} \,|\, \Psi_{n\bm{k}} }
            = & \,
                \sum_{\mu,\nu}^{N_{\rm basis}}
                C^{\bm{k}}_{\mu n} C^{\bm{k}*}_{\nu n}
                \left[
                    \dipmat +  \mathcal{R}^{\alpha}_\mu  \mathcal{S}_{\nu\mu}^{\bm{k}}
                \right].
        \label{eq:psi-r-psi}
    \end{align}
    Here \dipmat is the Fourier transform of the real-space \textit{dipole matrix} $\dipmatR$, defined as
    \begin{align}
        \dipmat
            = & \,
            \sum_{\bm{L}} e^{i \bm{k} \cdot \bm{L}} \,
            \underbrace{
                \braket{ \varphi_{\nu}^{\bm{0}} \,|\,  r^{\alpha} - \mathcal{R}^{\alpha}_\mu - L^{\alpha}  \,|\, \varphi_{\mu}^{\bm{L}} }
                }_{\dipmatR}.  \label{eq:D}
    \end{align}
    As detailed in the supplementary material, Sec.~S3, this formulation is convenient for implementation because it allows us to reuse the integration routines~\cite{Havu} employed to determine the translationally invariant elements of the Hamiltonian and overlap matrices.

    The second term, \ie the expectation value of the derivative with respect to the crystal momentum, is expanded similarly (see Appendix~\ref{sec:berry}), leading to
    \begin{align}
        \braket{ \Psi_{n\bm{k}}  \,|\,  \frac{\partial \Psi_{n\bm{k}}  }{\partial  k_\alpha } }
        =
        \sum_{\mu,\nu}^{N_{\rm basis}}
        C^{\bm{k}*}_{\nu n}
        \frac{ \partial C^{\bm{k}}_{\mu n} }{\partial k_\alpha }
        \mathcal{S}_{\nu\mu}^{\bm{k}}. \label{eq:psi-der-k}
    \end{align}
    The final expression for the Berry connection in Eq.~\eqref{eq:berry-connection-psi} in localized orbitals can therefore be written as the sum of a dipole term, $\dipoleterm$, and a Berry term, $\berryterm$:
    \begin{align}
        \mathcal{A}^\alpha_{n}\left(\bm{k}\right)
        &=
        \underbrace{
            \sum_{\mu,\nu}^{N_{\rm basis}}
            C^{\bm{k}*}_{\nu n}
            C^{\bm{k}}_{\mu n}
            \left[
                \dipmat +  \mathcal{R}^\alpha_\mu  \mathcal{S}_{\nu\mu}^{\bm{k}}
            \right]
        }_{\mathfrak{D}^\alpha_n(\bm{k})}
        + \notag \\
        & \quad
        \underbrace{
            i
            \sum_{\mu,\nu}^{N_{\rm basis}}
            C^{\bm{k}*}_{\nu n}
            \mathcal{S}_{\nu\mu}^{\bm{k}}
            \frac{ \partial C^{\bm{k}}_{\mu n} }{\partial k_\alpha }
        }_{\mathfrak{B}^\alpha_n(\bm{k})}.
        \label{eq:berry-connection-NAO}
    \end{align}

    \subsection{Numerical approximations}\label{sec:numerical}

    The expressions derived above are exact, but evaluating the polarization requires computing the reciprocal-space integrals given in Eqs.~\eqref{eq:tot-berry-phase} and~\eqref{eq:berry-phase}.
    In numerical calculations, these integrals must be replaced by finite sums over a \kpoint mesh.
    The dipole term $\dipoleterm$ can be evaluated directly, whereas the Berry term $\berryterm$ contains a gauge-dependent \kpoint-derivative that must be carefully discretized.
    Following Ref.~\cite{Resta2022SinglePoint}, we first recast the Berry term in Eq.~\eqref{eq:berry-connection-NAO} in terms of coefficients evaluated at $\bm{k}+q\bm{b}_\alpha$, where $q\in[0,1)$ is a continuous dimensionless variable.
    Taking the trace over the occupied-state indices $n$ and $m$ gives the following expression, where the derivative is evaluated at $q=0$:
    \begin{align}
        \sum_n^{\rm occ}\mathfrak{B}_n^\alpha(\bm{k})
            ={}&
            i \operatorname{Tr}\Biggl[
                \frac{\partial}{\partial q} \left(
                \sum_{\mu,\nu}^{N_{\rm basis}}
                C_{\nu n}^{\bm{k}*}
                \mathcal{S}_{\nu\mu}^{\bm{k}}
                C_{\mu m}^{\bm{k}+q\bm{b}_\alpha}\right)%
            \Biggr].
        \label{eq:berry-displaced-coefficients}
    \end{align}
    Given an invertible matrix $A$, the Jacobi formula from matrix calculus~\cite{magnus2019matrix} states that
    \begin{align}
        \frac{\partial}{\partial q} \ln \det(A) = \operatorname{Tr}\Bigl[A^{-1}\frac{\partial A}{\partial q}\Bigr],
        \label{eq:jacobi-formula}
    \end{align}
    which can be applied to our case since the matrix inside the trace in Eq.~\eqref{eq:berry-displaced-coefficients} reduces to the identity at $q=0$ given the orthonormality of the occupied \gls{KS} states~\cite{Resta2022SinglePoint}.
    Together with the fact that the Berry connection is real, this gives
    \begin{align}
        &\sum_n^{\rm occ}\mathfrak{B}_n^\alpha(\bm{k}) =
        \label{eq:jacobi-berry-displaced-coefficients} \\
        & \qquad = -
            \frac{\partial}{\partial q}
            \operatorname{Im}\ln\det\Biggl[
            \sum_{\mu,\nu}^{N_{\rm basis}}
            C_{\nu n}^{\bm{k}*}
            \mathcal{S}_{\nu\mu}^{\bm{k}}
            C_{\mu m}^{\bm{k}+q\bm{b}_\alpha}
            \Biggr]. \notag
    \end{align}

    For its numerical approximation, we represent each closed string \myloop by $N_{k_\alpha}$ equally spaced wave vectors $\bm{k}_1,\bm{k}_2,\ldots,\bm{k}_{N_{k_\alpha}}$ with the cyclic convention $\bm{k}_{N_{k_\alpha}+1}\equiv\bm{k}_1$, corresponding to a spacing of $\Delta k = 1/N_{k_\alpha}$ per string in the $[0,1)$ interval.
    Performing a Taylor expansion of Eq.~\eqref{eq:jacobi-berry-displaced-coefficients} around $q=0$, it is possible to show that the imaginary part of the zeroth-order term vanishes, and dividing the first-order term by the increment $\Delta k$ gives
    \begin{align}
        &\sum_n^{\rm occ}\mathfrak{B}_n^\alpha(\bm{k}_j) =
        \label{eq:jacobi-berry-finite-difference} \\
        & \qquad = -(\Delta k)^{-1}
            \operatorname{Im}\ln\det\Biggl[
            \sum_{\mu,\nu}^{N_{\rm basis}}
            C_{\nu n}^{\bm{k}_j*}
            \mathcal{S}_{\nu\mu}^{\bm{k}_j}
            C_{\mu m}^{\bm{k}_{j+1}}
            \Biggr] \notag.
    \end{align}
    We see from Eq.~\eqref{eq:jacobi-berry-finite-difference} that, within the finite-difference approximation, the Berry term associated with $\bm{k}_j$ actually depends on the \gls{KS} eigenvectors at a pair of neighboring \kpoints, $(\bm{k}_j,\bm{k}_{j+1})$, along the string \myloop.
    In the limit $\Delta k \rightarrow 0$, or equivalently $N_{k_\alpha}\rightarrow\infty$, Eq.~\eqref{eq:jacobi-berry-finite-difference} converges to the continuous expression in Eq.~\eqref{eq:jacobi-berry-displaced-coefficients}~\cite{Resta2022SinglePoint}.

    We next show that the finite-difference approximation introduced above preserves the gauge invariance of the Berry phase.
    We now return to Eq.~\eqref{eq:berry-phase}, which expresses the Berry phase entering the polarization as a line integral along a closed string \myloop.
    On the discrete mesh, the Berry-term contribution requires summing Eq.~\eqref{eq:jacobi-berry-finite-difference} over the $N_{k_\alpha}$ points of the string.
    Using the fact that a sum of logarithms is the logarithm of a product gives
    \begin{align}
        &
        \sum_{j=1}^{N_{k_\alpha}}
        \sum_n^{\rm occ}\mathfrak{B}_n^\alpha(\bm{k}_j) =
        \label{eq:sum-prod} \\
        & \qquad =
        -(\Delta k)^{-1}
        \operatorname{Im}\ln\Biggl[
        \prod_{j=1}^{N_{k_\alpha}}
        \det\Biggl(
            \sum_{\mu,\nu}^{N_{\rm basis}}
            C_{\nu n}^{\bm{k}_j*}
            \mathcal{S}_{\nu\mu}^{\bm{k}_j}
            C_{\mu m}^{\bm{k}_{j+1}}
        \Biggr)
        \Biggr]. \notag
    \end{align}
    As discussed in Sec.~\ref{subsec:MTP}, the \gls{KS} eigenvectors, and therefore their coefficients, can be multiplied by an arbitrary $\bm{k}$-dependent phase factor without changing their physical content.
    This issue is particularly relevant in numerical calculations, because independent diagonalizations do not guarantee that the \gls{KS} eigenvectors vary smoothly between \kpoints across reciprocal space~\cite{Resta2022SinglePoint}.
    The effect of this arbitrariness  can be represented in Eq.~\eqref{eq:sum-prod} by $\bm{k}$-dependent unitary transformations $\mathcal{U}^{\bm{k}_j}$, which account for these \textit{arbitrary phases}, as follows:
    \begin{align}
        &\Biggl[
            \sum_{\mu,\nu}^{N_{\rm basis}}
            C_{\nu n}^{\bm{k}_j*}
            \mathcal{S}_{\nu\mu}^{\bm{k}_j}
            C_{\mu m}^{\bm{k}_{j+1}}
        \Biggr]
        \longrightarrow
        \label{eq:cross-k-gauge-transformation}\\
        &\quad
        \left(\mathcal{U}^{\bm{k}_j}\right)^\dagger
        \Biggl[
            \sum_{\mu,\nu}^{N_{\rm basis}}
            C_{\nu n}^{\bm{k}_j*}
            \mathcal{S}_{\nu\mu}^{\bm{k}_j}
            C_{\mu m}^{\bm{k}_{j+1}}
        \Biggr]
        \mathcal{U}^{\bm{k}_{j+1}}.
        \notag
    \end{align}
    Using the multiplicativity of the determinant and the unitarity of $\mathcal{U}^{\bm{k}_{j}}$, the determinant factors contributed by two consecutive \kpoint pairs can be combined as
    \begin{align}
        &
        \det\left[\left(\mathcal{U}^{\bm{k}_{j}}\right)^\dagger\mathcal{U}^{\bm{k}_{j}}\right]
        = 1.
        \label{eq:unitary-determinant-cancellation}
    \end{align}
    Because the product in Eq.~\eqref{eq:sum-prod} runs over the closed loop, every unitary matrix is paired with its adjoint, while the cyclic condition $\bm{k}_{N_{k_\alpha}+1}\equiv\bm{k}_1$ closes the final pair.
    Consequently, Eq.~\eqref{eq:sum-prod} and the discretized polarization remains gauge invariant~\cite{Resta2022SinglePoint}.

    \section{Software Implementation and Parallelization}\label{sec:practical}

    This section describes the practical aspects of the implementation of the polarization formalism derived in the previous section in \aims.
    The scalable implementation presented here, based on \scalapack~\cite{scalapack}, builds on the earlier \lapack-based~\cite{lapack} routines described in Sec.~5.3 of Ref.~\cite{FHIaimsroadmap2025}.
    The generalized eigenvalue problems arising in the \scf cycle and during Fourier interpolation are solved through the \gls{ELSI} interface~\cite{ELSI}, which provides \aims access to serial and distributed solver libraries.
    Unless stated otherwise, the first three subsections discuss code-design choices shared by both implementations, whereas the final subsection compares their workflows and highlights their differences.

    \subsection{LU factorization}\label{sec:lu}

    In the \lapack implementation, the determinant in Eq.~\eqref{eq:jacobi-berry-finite-difference} is computed as the product of the eigenvalues of the occupied-space matrix inside the square brackets.
    The \scalapack implementation uses a LU factorization instead, which is substantially faster than a general eigendecomposition for distributed matrices.
    Denoting this matrix by $\mathsf{X}$, the factorization has the form $\mathsf{X}=\mathsf{P L U}$, where $\mathsf{P}$ is a permutation matrix, $\mathsf{L}$ is lower triangular with unit diagonal elements, and $\mathsf{U}$ is upper triangular~\cite{press2007numerical}.
    Its determinant is $\det\mathsf{X}=(-1)^s\prod_n \mathsf{U}_{nn}$ because $\det\mathsf{L}=1$ and $\det\mathsf{P}=(-1)^s$, where $s$ is the number of row interchanges performed during the factorization.
    Therefore, Eq.~\eqref{eq:jacobi-berry-finite-difference} reduces to
    \begin{align}
        & \sum_n^{\rm occ}\mathfrak{B}_n^{\alpha}(\bm{k}_j)
            = \label{eq:final-berry-finite-difference} \\
            & \notag \quad =
                -(\Delta k)^{-1}
                \left[
                    s(\bm{k}_j)\pi
                    +
                    \sum_n^{\rm occ}\operatorname{Im}\ln \mathsf{U}_{nn}(\bm{k}_j)
                \right].
    \end{align}

    Combining this result with the dipole term gives the electronic polarization implemented in \aims:
    \begin{align}
        & P^{\alpha}_e
        = \,
        -\frac{e\,|\bm{a}_\alpha|}{2\pi\,\Omega}\,
        \frac{1}{N_{\bm{k}_{\perp,\alpha}}\,N_{k_\alpha}}
        \sum_{\bm{k}_{\perp,\alpha}}
        \sum_{j=1}^{N_{k_\alpha}} \label{eq:Pe-numerical}  \\
        &
        \Biggl[
        \sum_n^{\rm occ}
        \mathfrak{D}_n^{\alpha}(\bm{k}_j)
        -
        (\Delta k)^{-1}
        \left(
        s(\bm{k}_j)\pi
        +
        \sum_n^{\rm occ} {\rm Im}\ln \mathsf{U}_{nn}(\bm{k}_j)
        \right)
        \Biggr],\notag
    \end{align}
    where $N_{\bm{k}_{\perp,\alpha}}$ and $N_{k_{\alpha}}$ are the numbers of \kpoints in the plane $\plane$ and the string \myloop, respectively.

    \subsection{Spin degrees of freedom}
\label{SecSpin}

    The derivations presented so far have neglected electronic spin.
    To include spin in the calculations, we augment each basis function with a spin index $\sigma=\pm1/2$:
    \begin{align}
        \varphi_{\mu}^{\bm{L}}(\bm{r}), \, \chi_{\mu}^{\bm{k}}(\bm{r}) \quad \longrightarrow \quad
        \varphi_{\mu}^{\sigma\bm{L}}(\bm{r}), \, \chi_{\mu}^{\sigma\bm{k}}(\bm{r}).
    \end{align}
    Within the standard one- and two-component framework, the present polarization implementation supports the following three spin treatments:%
    \begin{itemize}
        \item spin-unpolarized calculations: the Hamiltonian is spin independent and the two spin channels are identical;
        \item collinear spin-polarized calculations: the Hamiltonian is block diagonal in spin, with spin-dependent eigenvalues and eigenstates; $\sigma$ remains a good quantum number;
        \item \gls{soc}: the interaction is included perturbatively in the Hamiltonian in a non-self-consistent fashion, and the eigenstates become spinors that mix the two spin channels~\cite{Huhn2017_SOC}.
    \end{itemize}

    The overlap and dipole matrices depend only on the orbital part of the basis functions and are therefore diagonal in spin: $\mathcal{S}_{\nu\mu}^{\sigma'\sigma\bm{k}}=\delta_{\sigma'\sigma}\mathcal{S}_{\nu\mu}^{\bm{k}}$ and $\bm{\mathcal{D}}_{\nu\mu}^{\sigma'\sigma\bm{k}}=\delta_{\sigma'\sigma}\bm{\mathcal{D}}_{\nu\mu}^{\bm{k}}$.
    Accordingly, the electronic polarization $P^{\alpha}_e$ in Eq.~\eqref{eq:Pe-numerical} is evaluated as follows:
    \begin{itemize}
        \item spin-unpolarized: Eq.~\eqref{eq:Pe-numerical} is evaluated for one spin channel and multiplied by two;
        \item collinear spin-polarized: the contributions from the two channels are evaluated separately and summed;
        \item \gls{soc}: the basis-function indices $\mu$ and $\nu$ become composite spin-orbital indices.
    \end{itemize}

    Finally, let us note that \aims also provides a separate quasi-four-component relativistic framework~\cite{4c2,4c1}, which is, however, not yet supported within Berry-phase polarization calculations.

    \subsection{Additional numerical strategies}\label{subsec:periodic-gauge}

    The implementation employs additional numerical strategies to reduce the computational cost and enforce the appropriate boundary conditions.
    To evaluate the Berry term on a dense \kpoint grid without the corresponding cost of the self-consistent calculation, \aims converges the charge density on a coarser grid and Fourier-interpolates the wavefunction coefficients onto the denser grid~\cite{PickettFourier,JingkaiFourier}.
    We refer to these grids as the \scf and \nscf $\bm{k}$-grids, respectively.

    All user-specified \kpoints along each string are included in the Fourier interpolation.
    However, the interpolated coefficients require particular care at the ends of each string to satisfy the periodic gauge condition defined in Eq.~\eqref{eq:periodic-gauge}.
    The endpoints of each string \myloop differ by the reciprocal lattice vector $\bm{b}_\alpha$, and Fourier interpolation may assign them an arbitrary relative phase.
    Therefore, to impose the periodic gauge, the coefficients at the initial point, stored throughout the calculations, are reused at the final point to evaluate the Berry-term contribution associated with the last \kpoint pair in Eq.~\eqref{eq:Pe-numerical}.
    Consequently, $N_{k_\alpha}$ in Eq.~\eqref{eq:Pe-numerical} corresponds to the number of \nscf \kpoints specified by the user along the string minus one.
    The complete set of $N_{k_\alpha}+1$ interpolated points is given by $k_{\alpha,n}=-0.5+(n-1)/N_{k_\alpha}$, with $n=1,\ldots,N_{k_\alpha}+1$, while the $N_{k_\alpha}$ Berry-phase contributions are evaluated from consecutive pairs of these points.
    This construction is depicted in the following diagram.
    \begin{equation*}
    \includegraphics[width=0.8\linewidth]{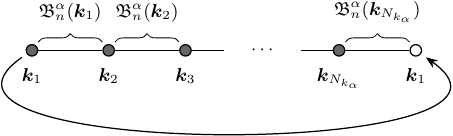}
\end{equation*}

    \begin{figure}[!t]
    \centering
    \includegraphics[width=\linewidth]{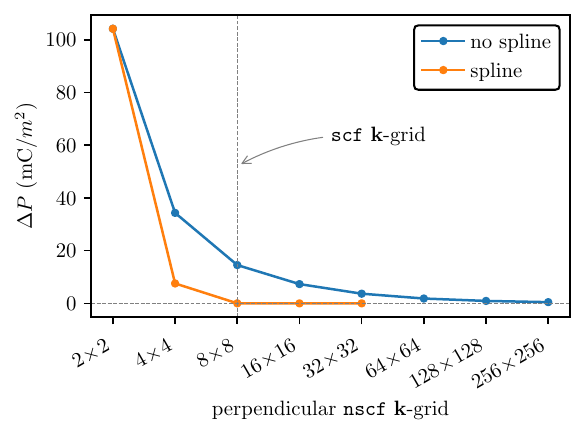}
    \caption{Absolute deviation of the polarization along the $[111]$ direction of rhombohedral \bto from the reference value as a function of the perpendicular \nscf \kpoint mesh, with and without spline interpolation of the Berry phases.
    The dashed vertical line indicates the \kpoint mesh density used in the self-consistent calculation.
    }
    \label{fig:spline}
\end{figure}

    Finally, to improve convergence with respect to the \kpoint mesh density, \aims uses a two-dimensional spline to interpolate the Berry phases $\phi_n(\bm{k}_{\perp,\alpha})$ over the plane $\plane$ before evaluating the total Berry phase $\phi_n^\alpha$ in Eq.~\eqref{eq:tot-berry-phase}.
    Because each Berry phase is defined modulo $2\pi$, the phases are first unwrapped by adding or subtracting integer multiples of $2\pi$ to remove discontinuities between neighboring $\bm{k}_{\perp,\alpha}$-points.
    The effect of this interpolation on the convergence of the polarization is illustrated in \fig{fig:spline} for \bto in the rhombohedral $R3m$ phase.
    The calculations were performed using the \texttt{PBEsol} \gls{xc} functional and the \texttt{tight} species defaults of \aims.
    The electronic density was converged using a \supercell{8} \scf \kpoint  mesh, while the Berry-phase strings were discretized with 80 \nscf \kpoints along the direction parallel to each reciprocal lattice vector.
    The density of the perpendicular \nscf mesh, which defines the number of strings, was then systematically increased while keeping the other computational parameters fixed.
    The polarization along the $[111]$ direction is shown as its absolute deviation $\Delta P$ from the result obtained with the densest spline-interpolated mesh.
    Without spline interpolation, the polarization converges only gradually with increasing \nscf $k_\perp$ points and remains noticeably different from the reference value when the perpendicular mesh density equals that of the underlying self-consistent calculation.
    In contrast, spline interpolation yields an essentially converged polarization already at $k_\perp=8$.
    We remind that the $k_\perp=8$ \nscf mesh includes both equivalent zone-boundary points and thus has only seven distinct points per perpendicular direction, making it slightly coarser than the $8\times8\times8$ \scf mesh.
    This demonstrates that the interpolation substantially reduces the number of explicitly evaluated Berry-phase strings required to converge the Brillouin-zone average, avoiding the need for very dense perpendicular \nscf \kpoint meshes.

    \begin{figure*}[!t]
    \centering
    \includegraphics[width=0.9\textwidth]{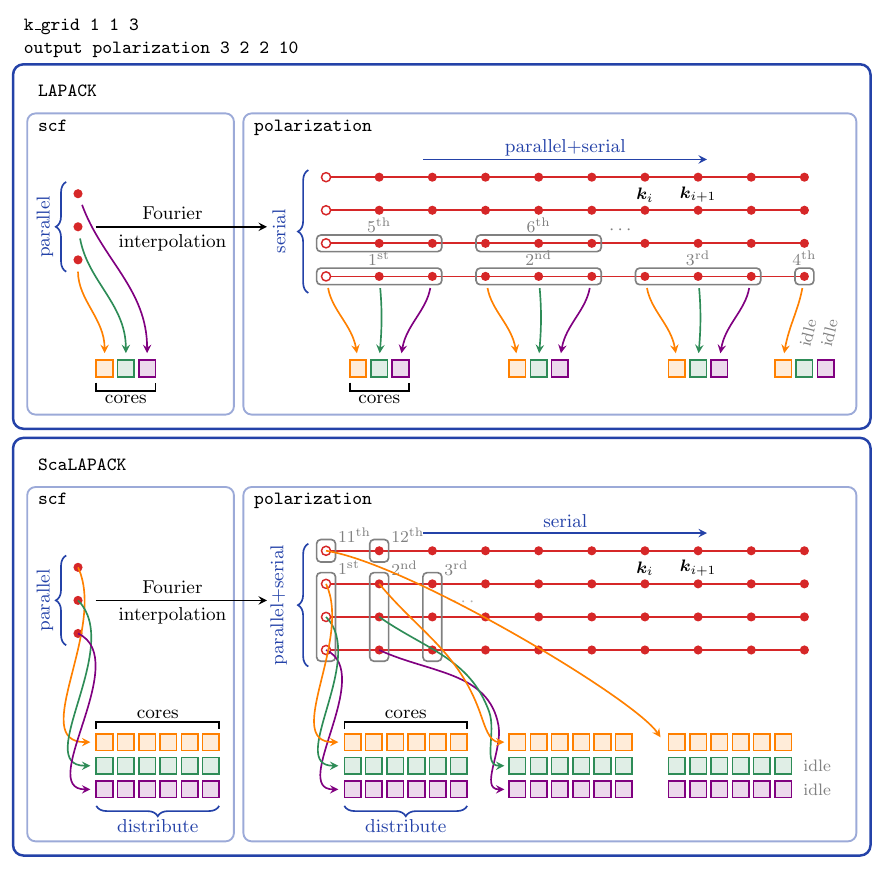}
    \caption[\lapack and \scalapack polarization workflows.]{
    Schematic comparison of the \lapack- and \scalapack-based workflows implemented in \aims for calculating the polarization.
    The inputs \texttt{k\_grid 1 1 3} and \texttt{output polarization 3 2 2 10} respectively define the \scf $\bm{k}$-grid ($1\times1\times3$) and request a polarization calculation along $\bm{a}_3$ on a $2\times2\times10$ \nscf $\bm{k}$-grid.
    These \scf \kpoints are shown in red in the left-hand \scf subpanels, while the \nscf grid comprises four strings of ten \kpoints each, as also shown in \fig{fig:strings}.
    In the polarization subpanels, gray rounded boxes and ordinal labels mark successive batches of calculations, which are processed one after another.
    In both cases, the gray \textit{idle} labels mark cores that remain idle because the final batch is incomplete.
    }
    \label{fig:scalapack}
\end{figure*}

\subsection{Parallelization and memory distribution}

    As already mentioned, \aims provides two implementations of the polarization formalism: one based on \lapack~\cite{FHIaimsroadmap2025} and one based on \scalapack.
    They use the same mathematical expressions and code-design choices described above, but differ in how the work and the required matrices are distributed over the available cores.
    This distribution is first established during the self-consistent cycle.
    When fewer cores than \scf \kpoints are available, \aims uses \lapack and distributes the \kpoints over the cores, so that each core is responsible for at least one \kpoint.
    This approach requires little communication and is efficient for small systems with many \kpoints, but each core must store the complete matrices required for its \kpoints, which limits the size of the systems that can be treated due to, \eg memory constraints.
    When the number of available cores exceeds the number of \scf \kpoints, a parallelization over \kpoints alone becomes less efficient, since some cores are left idling; \scalapack allows to leverage all available cores, since several cores can work together on the same \kpoint.
    The corresponding matrices are distributed over these cores, thereby reducing the memory required on each core and enabling calculations for larger systems.

    The polarization calculation is performed as a post-processing step after the self-consistent cycle and does not create a new parallelization or matrix distribution based on the \nscf grid, but repurposes the setup used during the self-consistent cycle.
    The \nscf grid hence only defines the polarization \kpoints and their arrangement into strings.
    In the \scalapack case, the self-consistent cycle establishes the groups of cores and their \gls{BLACS} grids.
    These groups are reused without any changes during post-processing, and their number determines how many polarization strings can be processed concurrently.
    \fig{fig:scalapack} shows how this inherited setup is used in the \lapack and \scalapack workflows.

    In the \lapack case, each core is responsible for one or more polarization \kpoints, so that the strings \myloop are processed sequentially while the \kpoints within each string are parallelized across the available cores.
    The upper panel of \fig{fig:scalapack} shows this ordering for a calculation using a $1\times1\times3$ \scf \kpoint grid and a $2\times2\times10$ \nscf grid on three cores.
    In this example, the ten \kpoints of the first string are processed in four successive batches of $3+3+3+1$.
    These batches are indicated by the numbered gray boxes and, after the first string is complete, the same procedure continues with the next string, whose first batch is marked as the fifth step in the figure.
    With this distribution, the \gls{KS} coefficients at the neighboring points $\bm{k}_{i-1}$ and $\bm{k}_i$ generally reside on different cores.
    Since both sets of coefficients are required for the Berry term in Eq.~\eqref{eq:jacobi-berry-finite-difference}, they must be exchanged through \gls{mpi} communication~\cite{mpi}.

    In the \scalapack case, several cores are responsible for each selected \kpoint and form a \gls{BLACS} grid, namely a two-dimensional arrangement used to distribute matrix blocks and coordinate communication.
    In the illustrated example of \fig{fig:scalapack}, we consider the same \scf and \nscf grids as previously, but now with 18 cores available.
    The 18 cores are divided into three groups of six, each forming a six-core \gls{BLACS} grid that operates on the matrices associated with one of the three selected \kpoints.
    The linear-algebra operations are performed directly on the distributed matrices, without reconstructing or storing them in full on an individual core.
    The polarization step therefore retains the memory distribution of the self-consistent cycle and does not introduce a separate memory bottleneck.
    Unlike the \lapack workflow, the \scalapack workflow processes the strings \myloop in parallel while traversing the \kpoints within each string sequentially.
    The four strings are therefore divided into a first batch of three strings and a second batch containing the remaining string.
    The code first traverses all ten \kpoint positions of the three strings in the first batch and only afterward traverses the fourth string, whose first two iterations are therefore marked $11^{\rm th}$ and $12^{\rm th}$.
    This ordering keeps the coefficients at $\bm{k}_{i-1}$ and $\bm{k}_i$ on the same group of cores, where they are retained until the next \kpoint is processed.
    It therefore avoids the additional \gls{mpi} communication step required by the \lapack implementation.

    \section{Validation and Benchmarks}\label{sec:results}

    This section assesses the correctness and performance of the distributed \scalapack implementation.
    We first test its numerical and physical reliability for polarization and \glspl{BEC} and then compare its computational cost with that of the original \lapack implementation.
    Finally, we examine the scaling of the \scalapack implementation with system size and its parallel scalability with core count using local and hybrid functionals.

    \subsection{Validation tests}

    \begin{table*}[!t]
    \centering
    \begin{tabular}[t]{@{} l c l l | r r @{}}
            \toprule
            system & direction & & xc & $\Delta P^3$ & $|\Delta\bm{P}|$  \\
            \midrule
            \bto & $[001]$   & this work                        & \texttt{LDA}    &       & 0.265 \\
            (tetragonal)            &           &                      & \texttt{PBEsol} &       & 0.345 \\
                        &           &                                  & \texttt{PBE}    &       & 0.473      \\
                        &           &                                  & \texttt{HSE06}  &       & 0.461 \\
            \myline{3-6}
                        &           & Ref.~\cite{choi2025first}       & \texttt{LDA}    &        & 0.260 \\
                        &           &                                  & \texttt{PBEsol} &       & 0.350 \\
                        &           &                                  & \texttt{HSE06}  &       & 0.450 \\
                        &           & Ref.~\cite{zhang2017comparative}& \texttt{LDA}    &        & 0.243 \\
                        &           &                                  & \texttt{PBE}    &        & 0.470 \\
                        &           &                                  & \texttt{HSE03}    &        & 0.407 \\
                \myline{3-6}
                        &           & Ref.~\cite{imura2014polarization}& Exp. &        & 0.273 \\
                        &           & Ref.~\cite{wieder1955electrical}\textsuperscript{\(\dagger\)} & Exp. &        & 0.260 \\
            \midrule
            \bto            & $[111]$   & this work             & \texttt{LDA}    & 0.177  & 0.308 \\
                        (rhombohedral)           &           &         & \texttt{PBEsol} & 0.226  & 0.392 \\
                        &           &                                  & \texttt{PBE}    & 0.288  & 0.501 \\
                        &           &                                  & \texttt{HSE06}  & 0.278  & 0.483 \\
                        \myline{3-6}
                        &           & Ref.~\cite{goh2016structural}    & \texttt{PBE}    & 0.212  & 0.368 \\
                        &           & Ref.~\cite{seo2013polytypes}     & \texttt{LDA}    & 0.148  & 0.256 \\
                        &           &                                  & \texttt{PBE}    & 0.265  & 0.461 \\
            \myline{3-6}
                        &           & Ref.~\cite{wieder1955electrical}\textsuperscript{\(\dagger\)} & Exp. & \multicolumn{1}{|c}{-} & 0.335 \\
            \bottomrule
    \end{tabular}
    \hspace{1cm}
    \begin{tabular}[t]{@{} l c l l | r r @{}}
            \toprule
            system & direction & & xc & $\Delta P^3$ & $|\Delta\bm{P}|$  \\
            \midrule
            \ce{BiFeO3} & $[111]$   & this work                        & \texttt{LDA}    &  0.405 & 1.002 \\
                        &           &                                  & \texttt{PBEsol} &  0.404 & 1.003 \\
                        &           &                                  & \texttt{PBE}    &  0.436 & 1.059 \\
                        &           &                                  & \texttt{HSE06}  &  0.441 & 1.073 \\
                        \myline{3-6}
                        &           & Ref.~\cite{bfo2005}              & \texttt{LDA}    & 0.404  & 0.987 \\
                        &           & Ref.~\cite{zhang2017comparative} & \texttt{LDA}    & 0.405  & 0.989 \\
                        &           &                                  & \texttt{PBE}    & 0.425  & 1.048 \\
                        &           &                                  & \texttt{HSE03}    & 0.447  & 1.103 \\
                        \myline{3-6}
                        &           & Ref.~\cite{bfo2007}\textsuperscript{\(\dagger\)}              & Exp. & \multicolumn{1}{|c}{-} & 1.000 \\
                        &           & Ref.~\cite{li2004dramatically}   & Exp. & \multicolumn{1}{|c}{-} & 1.000 \\
            \midrule
            \ce{PbTiO3} &  $[001]$  & this work                        & \texttt{LDA}    &       & 0.810 \\
                        &           &                                  & \texttt{PBEsol} &       & 0.901 \\
                        &           &                                  & \texttt{PBE}    &       & 1.256 \\
                        &           &                                  & \texttt{HSE06}  &       & 1.247 \\
                        \myline{3-6}
                        &           & Ref.~\cite{saghi1999first}       & \texttt{PBE96}  &       & 0.880 \\
                        &           & Ref.~\cite{zhang2017comparative} & \texttt{LDA}    &       & 0.798 \\
                        &           &                                  & \texttt{PBE}    &       & 1.255 \\
                        &           &                                  & \texttt{HSE03}   &       & 1.144 \\
                        \myline{3-6}
                        &           & Ref.~\cite{stephanovich2023selfpolarization}\textsuperscript{\(\dagger\)}
                                                                & Exp. &       & 0.82(2) \\
                        &           & Ref.~\cite{pto2004}              & Exp. &       & 0.965 \\
            \bottomrule
    \end{tabular}
    \caption[Validation: polarization differences.]{
    Polarization differences (expressed in $\mathrm{C}/\mathrm{m}^2$) between the representative centrosymmetric and polar structures for three systems and different \gls{xc} functionals.
    Previous fully relaxed first-principles results and representative experimental values are included for comparison.
    A dash denotes a value not reported in the cited study.
    The lattice and internal-site parameters are reported in the supplementary material, Sec.~S4.
    References marked with \textsuperscript{\(\dagger\)} report polarization values but no structural parameters.
    The vector $\Delta\bm{P}$ is expressed in Cartesian components, while the direction column identifies its crystallographic direction.
    The quantity $\Delta P^3\equiv P^3_{\mathrm{polar}}-P^3_{\mathrm{ref}}$ is the difference between the third direct-lattice components of the polar and centrosymmetric reference structures, as defined in Sec.~\ref{subsec:MTP}.
    For rhombohedral \bto and \ce{BiFeO3}, $\Delta\bm{P}$ points along $[111]$, and symmetry makes the three lattice-direction components equal, so only $\Delta P^3$ is listed.
    For tetragonal \bto and \ce{PbTiO3}, $\Delta\bm{P}$ points along $[001]$, so $\Delta P^3$ is identical to $|\Delta\bm{P}|$ and the value is reported only once.
    }
    \label{table:validation}
\end{table*}

    The polarization benchmarks comprise spin-unpolarized \bto, spin-collinear \ce{BiFeO3}, and \ce{PbTiO3} including \gls{soc}.
    For each material, we compare a centrosymmetric reference phase with one or two polar phases and calculate the corresponding polarization difference.
    \tab{table:validation} summarizes the crystallographic direction of the Cartesian polarization-difference vector $\Delta\bm{P}$, its component $\Delta P^3$ along $\bm{a}_3/|\bm{a}_3|$, and its magnitude $|\Delta\bm{P}|$, together with previous first-principles results and representative experimental values.
    For each material, we used four \gls{xc} functionals:
    \begin{enumerate*}[label=(\roman*)]
        \item \texttt{LDA}, as parameterized in Ref.~\cite{LDA};
        \item \texttt{PBE}, as formulated in Ref.~\cite{PBE};
        \item \texttt{PBEsol}, as formulated in Ref.~\cite{PBEsol};
        \item the hybrid functional \texttt{HSE06}, with a screening parameter $\omega=0.11~a_0^{-1}$ ($a_0$ is the Bohr radius) and an exact-exchange mixing coefficient $\alpha=0.25$, as defined in Refs.~\cite{HSE}.
    \end{enumerate*}

    To ensure quantitative convergence of the underlying electronic structure~\cite{Carbogno.2022}, the self-consistent calculations used the \texttt{tight} species defaults provided by \aims, an $8\times8\times8$ \scf $\bm{k}$-point grid, and the atomic ZORA relativistic treatment~\cite{zora}.
    For each polarization component, the subsequent \nscf calculation used a grid 10 times denser along the corresponding reciprocal-space direction: $80\times 8\times 8$ for the polarization along $\bm{a}_1$, $8 \times 80 \times8$ for $\bm{a}_2$, and $8 \times 8 \times 80$ for $\bm{a}_3$.
    Each phase was relaxed independently for each functional.
    The minor differences among the polarization values in \tab{table:validation} arise mostly due to minor structural differences.
    Tables~S1--S5 in the supplementary material, Sec.~S4, report the relaxation details, structural parameters, and coordinate conventions.

    For every combination of material, phase, and functional, \scalapack reproduces the \lapack polarization to the last reported digit, confirming that the distributed implementation preserves the original numerical results.

    We now address each individual validation test.
    \bto provides two spin-unpolarized benchmarks.
    We compare its cubic phase (space group $Pm\bar{3}m$) with two polar phases: tetragonal $P4mm$, whose polarization lies along $[001]$, and rhombohedral $R3m$, whose polarization points along $[111]$.
    For both polar phases, the \texttt{LDA} polarization from this implementation is compatible with the reported experimental values, whereas \texttt{PBE}, \texttt{PBEsol}, and \texttt{HSE03}/\texttt{HSE06} generally yield more distorted relaxed structures and correspondingly larger polarization magnitudes, as already observed in previous theoretical studies~\cite{wieder1955electrical,imura2014polarization,btobecbis,zhang2017comparative,choi2025first,seo2013polytypes,goh2016structural}.
    In the tetragonal phase, this trend is particularly pronounced for standard \texttt{PBE}, which suffers from a well-known supertetragonality error that conventional hybrid functionals can inherit, while \texttt{PBEsol} shows a milder overestimation of the tetragonality~\cite{btobecbis,zhang2017comparative,choi2025first}.

    \begin{table*}[!ht]
    \centering
    \begin{tabular}[t]{@{}l l l l @{}}
        \toprule
        \multicolumn{4}{c}{\ce{MgO}} \\
        \arrayrulecolor{gray}
        \midrule[\lightrulewidth]
        \arrayrulecolor{black}
        xc              & Reference                  & $a$ (\AA) & $Z^*_{\ce{Mg}}$ \\
        \midrule
        \texttt{LDA}    & this work                  & 4.162     & 1.975 \\
                        & Ref.~\cite{rybin2023band}  & -         & 1.97  \\
                        & Ref.~\cite{mgobec}         & 4.240     & 1.96  \\
                        & Ref.~\cite{YAMAGUCHI2022}  & 4.159     & 1.97  \\
                        & Ref.~\cite{YAMAGUCHI2022}* & 4.159     & 1.98  \\
        \arrayrulecolor{gray}
        \midrule[\lightrulewidth]
        \arrayrulecolor{black}
        \texttt{PBE}    & this work                  & 4.255     & 2.001 \\
                        & Ref.~\cite{mgobec2}        & 4.283     & 1.99  \\
                        & Ref.~\cite{varmgo}         & 4.253     & 2.00  \\
        \arrayrulecolor{gray}
        \midrule[\lightrulewidth]
        \arrayrulecolor{black}
        \texttt{PBEsol} & this work                  & 4.216     & 1.986 \\
                        & Ref.~\cite{varmgo}         & 4.214    & 1.99  \\
        \bottomrule
    \end{tabular}
    \hspace{1cm}
    \begin{tabular}[t]{@{}lllllll@{}}
        \toprule
        \multicolumn{7}{c}{\bto} \\
        \arrayrulecolor{gray}
        \midrule[\lightrulewidth]
        \arrayrulecolor{black}
        xc & Reference & $a$ (\AA) &
        $Z^*_{\ce{Ba}}$ & $Z^*_{\ce{Ti}}$ &
        $Z^*_{\ce{O}\perp}$ & $Z^*_{\ce{O}\parallel}$ \\
        \midrule
        \texttt{LDA}    & this work                     & 3.935 & 2.79 & 7.37 & -2.18 & -5.81 \\
                        & Ref.~\cite{btobecbis}         & 3.958 & 2.76 & 7.11 & -2.12 & -5.65 \\
                        & Ref.~\cite{zhong1994giant}    & 3.945 & 2.75 & 7.16 & -2.11 & -5.69 \\
                        & Ref.~\cite{ghosez1995born}    & 3.940 & 2.77 & 7.25 & -2.15 & -5.71 \\
                        & Ref.~\cite{rybin2023band}     & 3.934 & 2.79 & 7.35 & -2.17 & -5.80 \\
        \arrayrulecolor{gray}
        \midrule[\lightrulewidth]
        \arrayrulecolor{black}
        \texttt{PBE}    & this work                     & 4.017 & 2.76 & 7.43 & -2.14 & -5.90 \\
                        & Ref.~\cite{btobecbis}         & 4.035 & 2.73 & 7.17 & -2.08 & -5.73 \\
                        & Ref.~\cite{rybin2023band}     & 4.017 & 2.76 & 7.41 & -2.14 & -5.89 \\
        \arrayrulecolor{gray}
        \midrule[\lightrulewidth]
        \arrayrulecolor{black}
        \texttt{PBEsol} & this work                     & 3.968 & 2.79 & 7.38 & -2.16 & -5.84 \\
                        & Ref.~\cite{rybin2023band}     & 3.967 & 2.78 & 7.36 & -2.16 & -5.83 \\
        \bottomrule
    \end{tabular}
    \caption{
    Born effective charges of \ce{MgO} and cubic \bto, in units of the elementary charge $e$, compared with previous first-principles results.
    In \ce{MgO}, the \gls{BEC} tensors of \ce{Mg} and \ce{O} are diagonal and isotropic, and the acoustic sum rule implies $Z^*_{\ce{O}}=-Z^*_{\ce{Mg}}$; therefore, only $Z^*_{\ce{Mg}}$ is reported.
    The asterisk (*) next to Ref.~\cite{YAMAGUCHI2022} indicates that the reported \gls{BEC} was evaluated using the finite-electric-field method implemented in that work.
    For \bto, only the symmetry-inequivalent components of the \gls{BEC} tensors are reported.
    A dash denotes a value not reported in the cited study.
    For cubic \bto, $Z^*_{\ce{Ba}}$ and $Z^*_{\ce{Ti}}$ are the isotropic diagonal components of the corresponding \gls{BEC} tensors, while $Z^*_{\ce{O}\parallel}$ and $Z^*_{\ce{O}\perp}$ are the oxygen components parallel and perpendicular to a Ti--O bond, respectively.}
    \label{table:bec}
\end{table*}

    \ce{BiFeO3} tests the spin-collinear implementation.
    This multiferroic exhibits coexisting antiferromagnetic and ferroelectric order~\cite{bfo2009}.
    We therefore compare its centrosymmetric $R\bar{3}c$ phase with its polar $R3c$ phase, both of which exhibit antiferromagnetic order.
    Although centrosymmetric, the $R\bar{3}c$ phase has a nonzero formal ``non-switchable'' polarization of $Q_\alpha/2$ along each periodic direction, arising solely from the ionic contribution, where $Q_\alpha$ is the polarization quantum defined previously~\cite{bfo2009}.
    This counterintuitive half-quantum polarization is compatible with centrosymmetry and has also been reported for other centrosymmetric systems~\cite{Resta2007}.
    Although this value is not reported in \tab{table:validation}, it can be inferred from the lattice parameters reported in Table~S3 and was included as the polarization of the centrosymmetric phase when calculating the polarization difference reported in \tab{table:validation}.
    The calculated components of the Cartesian polarization difference along the three lattice-vector directions are symmetry-equivalent because $\Delta\bm{P}$ points along $[111]$.
    The polarization magnitudes from this implementation agree with previous fully relaxed calculations from the cited studies and are also compatible with the experimental values listed in \tab{table:validation}.

    \ce{PbTiO3} tests the implementation for calculations including \gls{soc}.
    We compare its centrosymmetric cubic phase (space group $Pm\bar{3}m$) with its ferroelectric tetragonal phase (space group $P4mm$), whose polarization lies along $[001]$.
    As in \bto, the calculated polarization differences increase from \texttt{LDA} to \texttt{HSE06}, consistent with the progressively stronger distortion reported in Table~S4.
    Earlier first-principles calculations show the same substantial functional dependence~\cite{saghi1999first,zhang2017comparative}.
    The \texttt{LDA} and \texttt{PBEsol} results are compatible with the experimental single-crystal and film measurements listed in \tab{table:validation}, whereas \texttt{HSE06} yields a substantially larger polarization~\cite{stephanovich2023selfpolarization,pto2004}.

    The final test considers the \glspl{BEC} of \ce{MgO} and cubic \bto.
    We evaluate the \glspl{BEC} using central finite differences of the polarization, with atomic displacements of $\pm 10^{-3}$~\AA~along the symmetry-inequivalent directions.
    For each of the \gls{xc} functionals considered for comparison -- namely \texttt{LDA}, \texttt{PBE}, and \texttt{PBEsol} -- the unit cells were relaxed following the same procedure described above.
    All calculations used the \texttt{tight} species defaults, the atomic ZORA relativistic treatment, and a \supercell{16} \scf \kpoint-grid.
    The polarization was evaluated along each of the three periodic directions using \nscf \kpoint-grids of $64\times16\times16$ along $\bm{a}_1$ and the corresponding permutations along $\bm{a}_2$ and $\bm{a}_3$.
    As shown in \tab{table:bec}, the calculated \glspl{BEC} of both \ce{MgO} and \bto are in good agreement with the reported reference values.
    As in the polarization benchmarks, differences among the reported values arise mainly from differences in the adopted structures.
    Accordingly, the agreement among the \glspl{BEC} correlates with that among the relaxed lattice constants; this is especially evident for \bto, for which calculations with nearly identical lattice constants yield nearly identical \glspl{BEC}.
    The small residual differences may be attributed to differences in basis representations and core-electron treatments, as some of the reference studies employed \gls{DFT} codes other than \aims or used plane-wave pseudopotential approaches.
    Taken together, these comparisons provide an additional validation of the implementation.

    \subsection{\lapack vs \scalapack comparison}\label{sec:performance}

    Having established numerical agreement, we compare the CPU time and memory requirements of the \lapack and \scalapack implementations on a common \ce{MgO} benchmark.

    \begin{figure}[!tbp]
    \centering
    \includegraphics[width=\linewidth]{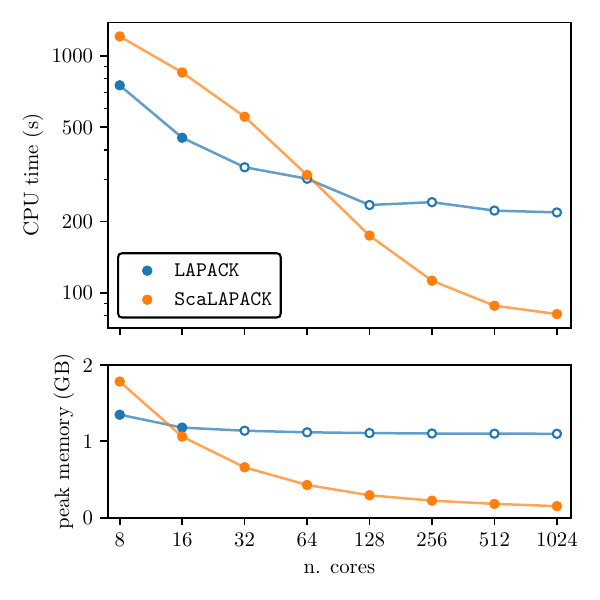}
    \caption{Comparison of CPU time and peak tracked memory per core for \lapack and \scalapack using a \supercell{4} supercell of \ce{MgO} (128 atoms).
    The \scf and \nscf $\bm{k}$-grids are $2\times2\times2$ and $2\times2\times16$, respectively, with the polarization evaluated along the \nth{3} lattice direction.
    The memory value is the maximum over all parallel tasks reported by \aims and corresponds to the peak tracked memory per core.
    }
    \label{fig:scalability}
\end{figure}

    \fig{fig:scalability} compares the computational time and memory requirements for a representative system: a \supercell{4} supercell of \ce{MgO} (128 atoms).
    In these simulations, we used the \texttt{light} species defaults, \texttt{LDA}, and a \supercell{2} \scf $\bm{k}$-grid, and computed the polarization along the \nth{3} periodic direction using a $2\times2\times16$ \nscf $\bm{k}$-grid.
    \fig{fig:scalability} reports both the net CPU time required to compute the polarization and the peak memory requirement.
    The reported net CPU time was evaluated as the difference between the CPU time of one \scf cycle followed by the polarization calculation and that of one \scf cycle without the polarization calculation.
    Both runs started from the same pre-converged charge density stored in binary files.
    The \lapack implementation shows a limited scalability beyond 128 cores.
    The precise core count at which the plateau begins depends on the system and computational parameters, but the plateau itself is intrinsic to the implementation.
    Although the plateau might be expected to begin as soon as the number of cores exceeds the number of \kpoints along the \nth{3} periodic direction (\ie 16), it appears only around 128 cores.
    \fig{fig:scalability} reports the \lapack data (empty circles) corresponding to the number of cores exceeding the number of \kpoints along the \nth{3} periodic direction, and therefore corresponding to a regime in which some cores are partially idle.
    This delayed onset arises because the construction of the real-space dipole matrix $\bm{\mathcal{D}}_{\nu\mu}^{\bm{L}}$ (see Eq.~\eqref{eq:D} and the supplementary material, Sec.~S3) is independent of the chosen linear-algebra library (\lapack or \scalapack).
    In fact, the dipole matrix is evaluated through the integration subroutines available in \aims, which distribute the workload over all available cores.
    This is illustrated in more detail in the supplementary material, Sec.~S5.

    Conversely, the \scalapack implementation remains scalable at larger core counts by distributing both the computational work and memory across substantially more cores.
    In this example, the computational costs of the two implementations are comparable up to 128 cores, whereas the \scalapack implementation becomes faster at higher core counts.
    The bottom panel shows the peak memory consumption during the dipole evaluation.
    The peak memory of the \lapack implementation saturates as the number of cores increases, indicating that its memory usage becomes effectively independent of the parallel decomposition beyond a certain core count, hence limiting the scalability for extended system sizes.
    In contrast, the peak memory of the \scalapack implementation decreases monotonically with the number of cores, reflecting the distribution of both computational work and memory across processes.

    \subsection{Performance, scalability, and scaling}

        We now discuss both the scaling of the polarization implementation with system size and its parallel scalability.
        Here, the former describes how the computational time changes with number of atoms at a fixed number of cores, whereas the latter describes how it changes with the number of cores for a fixed system.

        \begin{figure}[!t]
    \centering
    \includegraphics[width=\linewidth]{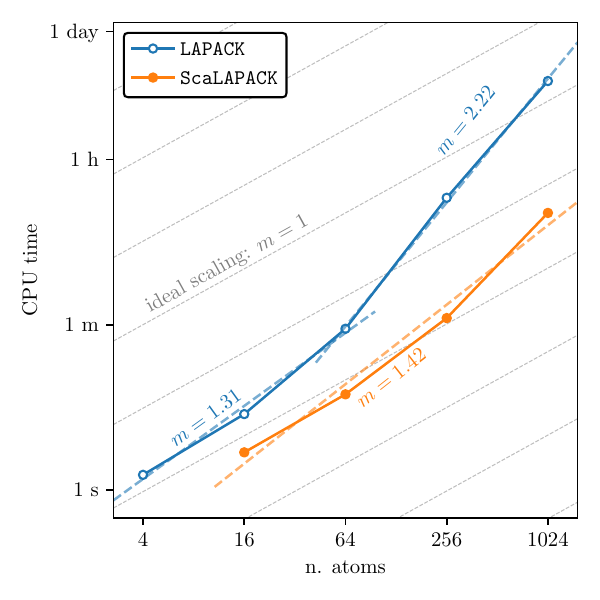}
    \caption{%
    Net CPU time for polarization in \ce{MgO} supercells using 512 cores with \lapack and \scalapack.
    The \scf $\bm{k}$-point density is kept constant across supercells, using an \scf $\bm{k}$-grid ranging from $8\times16\times16$ for the $2\times1\times1$ supercell with 4 atoms to $2\times2\times2$ for the $8\times8\times8$ supercell with 1024 atoms.
    The corresponding \nscf $\bm{k}$-grid is fourfold denser along the polarization direction, ranging from $32\times16\times16$ for the smallest supercell to $8\times2\times2$ for the largest.
    Dashed lines show least-squares power-law fits: a single fit for \scalapack and two separate fits for \lapack over the smaller and larger system-size ranges.
    }
    \label{fig:scalability-atoms}
\end{figure}

\begin{figure}[!tbp]
    \centering
    \includegraphics[width=\linewidth]{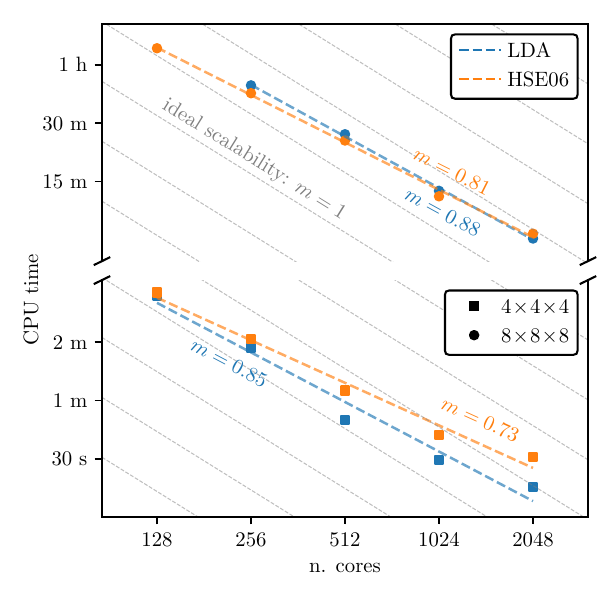}
    \caption{
    Net polarization time as a function of the number of cores for the $4\times4\times4$ and $8\times8\times8$ \bto supercells, calculated with \texttt{LDA} and \texttt{HSE06}.
    The used \scf and \nscf $\bm{k}$-grids are reported in \tab{table:simulation-parameters}.
    The broken y-axis omits a range without data; the gray dashed lines indicate ideal scalability ($m=1$), and the colored dashed lines are power-law fits.
    Blue and orange distinguish \texttt{LDA} and \texttt{HSE06}, respectively, while square and circular markers distinguish the $4\times4\times4$ and $8\times8\times8$ supercells, respectively.
    }
    \label{fig:performance}
\end{figure}

We first examine how the net polarization time changes with system size at a fixed allocation of 512 cores.
We benchmarked the \lapack and \scalapack implementations on \ce{MgO} supercells ranging from 4 to 1024 atoms, using \texttt{LDA} and the \texttt{light} species defaults.
We reduced the \scf \kpoint grid in inverse proportion to each supercell dimension to maintain the sampling density of a \supercell{16} grid for the two-atom primitive cell.
For polarization, the \nscf grid was four times as dense as the corresponding \scf grid along the polarization direction and had the same density in the two transverse directions.
Thus, the number of explicit \kpoints decreased as the supercells grew, while the sampling density remained fixed.

\fig{fig:scalability-atoms} shows the net polarization times and least-squares fits to $y=Ax^m$, where $x$ is the number of atoms.
A single power-law fit to the \scalapack timings gives $m=1.42$.
For \lapack, separate fits give $m=1.31$ for 4--64 atoms and $m=2.22$ for 64--1024 atoms.
\scalapack was faster at every size for which both timings are shown, and its speedup increased with system size.
At 1024 atoms, \lapack took about seven hours and \scalapack took about 16 minutes, a speedup of about a factor of 25.

As the \ce{MgO} supercells grow, fewer $\bm{k}$-points are needed to maintain the same sampling density.
In the \lapack implementation, some steps process one string at a time, with each core working on a different k-point in that string.
As the strings get shorter, fewer of the 512 allocated cores can work on those steps simultaneously.
The fitted exponents therefore describe the measured timings for this benchmark, which combines increasing system size with a decreasing number of explicit \kpoints.
The supplementary material, Sec.~S6 shows the main contributions to the \scalapack timings.

        \begin{table}[!t]
    \centering
    \begin{tabular}{@{}r c c c l l@{}}
        \toprule
        \text{atoms} & \text{supercell} & \scf $\bm{k}$-grid & \nscf $\bm{k}$-grid & xc & \text{$m$} \\
        \midrule
        320  & $4\times4\times4$ & $2\times2\times2$ & $2\times2\times4$ & \texttt{LDA}   & 0.85(8) \\
             &                   &                   &                           & \texttt{HSE06} & 0.73(5) \\
        \arrayrulecolor{gray}
        \midrule[\lightrulewidth]
        \arrayrulecolor{black}
        2560 & $8\times8\times8$ & $1\times1\times1$ & $1\times1\times2$ & \texttt{LDA}   & 0.88(2) \\
             &                   &                   &                           & \texttt{HSE06}& 0.81(2) \\
        \bottomrule
    \end{tabular}
    \caption{Simulation parameters and fitted core-count scalability exponents for the data shown in Fig.~\ref{fig:performance}.
    The \nscf $\bm{k}$-grid is used to calculate the polarization along the third lattice direction.}
    \label{table:simulation-parameters}
\end{table}

Finally, we assess the parallel scalability of the \scalapack implementation by monitoring the net polarization time at different core counts for two supercells of \bto.
We considered $4\times4\times4$ and $8\times8\times8$ supercells of the tetragonal five-atom unit cell, containing 320 and 2560 atoms, respectively.
Both supercells had spontaneous polarization along the third lattice direction.
We first computed the polarization using both \texttt{LDA} and the hybrid functional \texttt{HSE06}, with the \scf and \nscf $\bm{k}$-grids listed in \tab{table:simulation-parameters}.
In all cases, we used the \texttt{light} species defaults.
We repeated the calculations at different core counts between 128 and 2048.
We then fitted the net polarization time to a power law $y=Ax^{-m}$, where $x$ is the number of cores.
The results in the left panel of \fig{fig:performance} show good strong scaling, with fitted exponents below the ideal value of $m=1$.

The deviation from the ideal exponent $m=1$ is expected because, as the number of cores increases, the work assigned to each rank decreases while \gls{mpi} communication, synchronization, and distributed-matrix overheads become increasingly important.
The two functionals yield similar scalability exponents, indicating that the functional has only a limited effect on the strong scalability of the polarization calculation.
Still, slightly smaller scalability exponents are observed in the \texttt{HSE06} case, particularly for the smaller supercell.
To identify its origin, we performed a detailed timing analysis reported in the supplementary material, Sec.~S7.
Because of the high density of \nscf $k$-points in these simulations, the Fourier interpolation~\cite{PickettFourier,JingkaiFourier} is the dominant contribution to the net polarization time, since it includes the expensive diagonalization step.
The remaining considerable contributions, \ie the Berry and dipole terms and the dipole-matrix construction, are nearly independent of the functional.
Fourier interpolation involves the construction of interpolated Hamiltonian and overlap matrices followed by the solution of a distributed eigenvalue problem on each \nscf $k$-point.
For hybrid functionals, it additionally requires the construction of the $\bm{k}$-dependent Fock matrix, which entails additional distributed computation and, especially, \gls{mpi} communication.
The \gls{mpi} communication overhead becomes relatively more apparent at high core counts for the smaller supercell, as discussed in the supplementary material, Sec.~S7 and can be identified as the source of the small difference in performance for the hybrid functional.
Overall, these results show that the polarization calculation remains strongly scalable across functionals, with its modest functional dependence observed in smaller systems originating primarily from the Fourier-interpolation step.

This is further substantiated in the supplementary material, Sec.~S9.
There we repeated the analysis for both \bto supercells using deliberately dense $\bm{k}$-grids, thereby increasing the workload per core.
The obtained exponents are $m=0.94$ and $m=0.95$ for the smaller and larger supercells, respectively, and are therefore significantly closer to the ideal value of $m=1$.
A similar analysis, reported in Sec.~S8, shows how the performance of the code changes with the functional and box size for different liquid water boxes.
These results show that the scalability of the code is largely independent of the chosen functional at larger system sizes and that near-ideal scalability can be reached for the calculation of the polarization for sufficiently extended systems.

    \section{Conclusion}\label{sec:conclusion}

    In this work, we have presented a scalable implementation of the modern theory of polarization in the all-electron FHI-aims code, which brings Berry-phase polarization calculations to system sizes that have so far not been routinely accessible.
    This is enabled by two central aspects of the implementation.
    On the formal side, expressing the Berry connection in the NAO basis through a translationally invariant real-space dipole matrix and a finite-difference term built from Kohn--Sham coefficients at neighboring $\bm k$-points avoids $\bm k$-dependent phase-operator integrals.
    In turn, this allows to use the massively parallel real-space integration routines~\cite{Havu} already present in \aims.
    On the algorithmic side, which is the novel point of this paper, distributing the matrices of each $\bm k$-point with ScaLAPACK, while inheriting the matrix distribution of the self-consistent cycle, removes the memory bottleneck of the $\bm k$-point-parallel LAPACK implementation that was severely limiting the treatable system sizes.
    Overall, this results in pronounced computational speedups, \eg of about a factor of 25 for the 1,024 atom MgO supercell when using 512 cores.
    Concurrently, the code retains strong scalability up to 2,048 cores, as demonstrated for the 2,560-atoms BaTiO$_3$ supercell both at the local and hybrid level of theory.

    Being independent of the employed treatment for exchange-correlation, the implementation can be used with all functionals available in FHI-aims.
    We explicitly showed that the distributed implementation reproduces the LAPACK results for non-magnetic, collinear spin-polarized, and spin-orbit-coupled polar oxides, and yields polarizations and Born effective charges consistent with previous first-principles results for local and semilocal, but also with hybrid functionals.
    Polarizations and Born effective charges, and with them infrared intensities~\cite{Vibes} and electric-field coupling, thus become accessible at the hybrid-functional level for systems containing thousands of atoms.
    This makes it practical to generate large, accurate reference datasets for machine-learning models of polarization-related properties.

    Several developments could further optimize the method, including a more balanced distribution of strings and $\bm k$-points to reduce idle cores, the extension to related Berry-phase quantities such as Wannier centers, and the generalization to the quasi-four-component relativistic framework of \aims~\cite{FHIaimsroadmap2025}.
    Computationally, the Fourier interpolation onto the \texttt{nscf} $\bm k$-grid dominates the numerical cost at large system sizes or when a dense \texttt{nscf} grid is required (see the supplementary material, Sec.~S6).
    This is not too surprising, given that this step includes the eigenvalue problem solution, which formally scales cubically with system sizes.
    By building on the \gls{ELSI} library for electronic structure solvers~\cite{ELSI}, our implementation can directly benefit from any present and future advancements in this field~\cite{ELPA,ELPA2}.

    \section*{Supplementary Material}
        See the supplementary material for the derivation of the derivative with respect to the reciprocal-lattice coordinates (Sec.~S1), the reciprocal-space matrix elements (Sec.~S2), and the dipole matrix elements (Sec.~S3).
        It also reports the structural relaxations and geometries of the test systems (Sec.~S4), the timing components of the \lapack calculation (Sec.~S5), the timing profile of the polarization costs (Sec.~S6), and a component-resolved scalability analysis (Sec.~S7).
        Finally, it presents the parallel scalability for liquid-water polarization (Sec.~S8) and with denser $\bm{k}$-grids (Sec.~S9).

    \section*{Acknowledgments}
        E.S. thanks Christoph D{\"a}hn for useful discussions during the development of the code.
        E.S. acknowledges support from the UFAST International  Max Planck Research School.
        C.C. and M.R. thank Karsten Reuter for supporting E.S. on a research visit for this project.
        We acknowledge project support from the Max Planck Computing and Data Facility.

    \section*{Author Declarations}

    \subsection*{Conflict of Interest}
        The authors have no conflicts to disclose.

    \subsection*{Author Contributions}
        \textbf{Elia Stocco}: Data curation (lead); Formal analysis (equal); Investigation (lead); Software (equal); Validation (equal); Visualization (lead); Writing -- original draft (lead); Writing -- review \& editing (equal).
        \textbf{James A. Green}: Formal analysis (equal); Investigation (supporting); Methodology (equal); Resources (equal); Software (equal); Supervision (equal); Validation (equal); Writing -- review \& editing (equal).
        \textbf{Mariana Rossi}: Conceptualization (equal); Formal analysis (equal); Funding acquisition (lead); Methodology (equal); Project administration (equal); Resources (equal); Software (supporting); Supervision (equal); Validation (equal); Writing -- original draft (supporting); Writing -- review \& editing (equal).
        \textbf{Christian Carbogno}: Conceptualization (lead); Formal analysis (equal); Funding acquisition (supporting); Methodology (equal); Project administration (equal); Resources (equal); Software (equal); Supervision (equal); Validation (equal); Writing -- original draft (supporting); Writing -- review \& editing (equal).

    \section*{Data and Code Availability}

        The data that support the findings of this study are openly available in Zenodo at \url{https://doi.org/10.5281/zenodo.24537194}, Ref.~\cite{zenodo}.
        The associated actively maintained repository is available on GitHub at \url{https://github.com/EliaStocco/ScaLAPACK-polarization-FHI-aims}.
        The figures were created mainly using Matplotlib~\cite{matplotlib} and PGF/TikZ ~\cite{tantau2023tikz}.
        The implementation presented in this work is available as part of \aims; further information is provided on the official website at \url{https://fhi-aims.org/}.

    \appendix
        \section{Berry connection in the atomic gauge}\label{sec:berry-atomic}
    We now express the previous equations in a different gauge, \ie the atomic gauge~\cite{atomicgaugescipost}.
    We start by redefining the Bloch-like basis functions as
    \begin{align}
        \tilde{\chi}_{\mu}^{\bm{k}}(\bm{r}) = & \,
        e^{i \bm{k} \cdot    \bm{R}_\mu   } \,
        \chi_{\mu}^{\bm{k}}(\bm{r}). \label{eq:atomic-gauge-basis}
    \end{align}
    We can then solve the \gls{KS} equations in this new basis, which leads to the following generalized eigenvalue problem:
    \begin{align}
        \sum_{\mu}^{N_{\rm basis}} \,
        \tilde{\mathcal{H}}^{\bm{k}}_{\nu\mu}
        \,
        \tilde{C}_{\mu n}^{\bm{k}}
        = & \,
        \varepsilon_{n}^{\bm{k}}  \,
        \sum_{\mu}^{N_{\rm basis}} \,
        \tilde{\mathcal{S}}_{\nu\mu}^{\bm{k}}
        \,
        \tilde{C}_{\mu n}^{\bm{k}}
    \end{align}
    where the \gls{KS} coefficients $\tilde{C}_{\mu n}^{\bm{k}}$ and the matrix elements of any operator, such as $\tilde{\mathcal{H}}^{\bm{k}}_{\nu\mu}$ or $\tilde{\mathcal{S}}_{\nu\mu}^{\bm{k}}$, are related to their counterparts in the lattice gauge through the unitary transformation $U_{\mu\nu}^{\bm{k}}= e^{ i \bm{k} \cdot \bm{R}_\mu  }\delta_{\mu\nu}$ as follows:
    \begin{align}
        \tilde{C}_{\mu n}^{\bm{k}}
            = & \,
            e^{  - i \bm{k} \cdot \bm{R}_\mu  } \, C_{\mu n}^{\bm{k}} \label{eq:coeff-lattice-atomic} \\
        \tilde{A}^{\bm{k}}_{\nu\mu}
            = & \,
            e^{  i \bm{k} \cdot \left( \bm{R}_\mu - \bm{R}_\nu \right) } \, A^{\bm{k}}_{\nu\mu} \label{eq:operator-lattice-atomic}
    \end{align}
    Here, $\hat{A}$ represents any operator, such as the overlap, Hamiltonian, or dipole operator.

    Inserting Eqs.~\eqref{eq:coeff-lattice-atomic} and~\eqref{eq:operator-lattice-atomic} into the lattice-gauge Berry connection in Eq.~\eqref{eq:berry-connection-NAO}, with the operator transformation applied to the dipole and overlap matrices, gives the Berry connection in the atomic gauge:
    \begin{align}
        \mathcal{A}_n^\alpha(\bm{k})
            ={}&
            \underbrace{
                \sum_{\mu,\nu}^{N_{\rm basis}}
                \tilde{C}_{\nu n}^{\bm{k}*}
                \tilde{C}_{\mu n}^{\bm{k}}
                \tilde{\mathcal{D}}_{\nu\mu}^{\bm{k},\alpha}
            }_{\dipoleterm}
            \label{eq:appendix-berry-connection-atomic}\\
        &+
            \underbrace{
                i\sum_{\mu,\nu}^{N_{\rm basis}}
                \tilde{C}_{\nu n}^{\bm{k}*}
                \frac{\partial\tilde{C}_{\mu n}^{\bm{k}}}{\partial k_\alpha}
                \tilde{\mathcal{S}}_{\nu\mu}^{\bm{k}}
            }_{\berryterm}.
        \notag
    \end{align}

    In practice, \aims provides the lattice-gauge coefficients $C_{\mu n}^{\bm{k}}$, rather than the transformed coefficients $\tilde{C}_{\mu n}^{\bm{k}}$.
    Expressing the undifferentiated coefficients and the dipole and overlap matrices in Eq.~\eqref{eq:appendix-berry-connection-atomic} through Eqs.~\eqref{eq:coeff-lattice-atomic} and~\eqref{eq:operator-lattice-atomic} gives the equivalent form
    \begin{align}
        \mathcal{A}_n^\alpha(\bm{k})
            ={}&
            \sum_{\mu,\nu}^{N_{\rm basis}}
            C_{\nu n}^{\bm{k}*}
            C_{\mu n}^{\bm{k}}
            \mathcal{D}_{\nu\mu}^{\bm{k},\alpha}
            \label{eq:appendix-berry-connection-atomic-lattice-coefficients}\\
        &+
            i\sum_{\mu,\nu}^{N_{\rm basis}}
            C_{\nu n}^{\bm{k}*}
            \mathcal{S}_{\nu\mu}^{\bm{k}}
            e^{i\bm{k}\cdot\bm{R}_\mu}
            \frac{\partial\tilde{C}_{\mu n}^{\bm{k}}}{\partial k_\alpha}.
        \notag
    \end{align}

    Introducing the scalar displacement $q$ along $\bm{b}_\alpha$ and using Eq.~\eqref{eq:coeff-lattice-atomic}, the Berry term in Eq.~\eqref{eq:appendix-berry-connection-atomic-lattice-coefficients} can be written entirely in terms of the lattice-gauge coefficients available in \aims:
    \begin{align}
        \berryterm
        ={}&
        i\frac{\partial}{\partial q}
        \left[
            \sum_{\mu,\nu}^{N_{\rm basis}}
            C_{\nu n}^{\bm{k}*}\mathcal{S}_{\nu\mu}^{\bm{k}}
            e^{-iq\bm{b}_\alpha\cdot\bm{R}_\mu}
            C_{\mu n}^{\bm{k}+q\bm{b}_\alpha}
        \right]
        ,
        \label{eq:atomic-berry-q}
    \end{align}
    where the derivative is taken at $q=0$.Applying the occupied-state trace, Jacobi formula, and forward finite difference introduced in Sec.~\ref{sec:numerical} gives the atomic-gauge analogue of Eq.~\eqref{eq:jacobi-berry-finite-difference}:
    \begin{align}
        &\sum_n^{\rm occ}\mathfrak{B}_n^\alpha(\bm{k}_j)
        =\label{eq:atomic-berry-finite-difference}\\
        & =-(\Delta k)^{-1}
        \operatorname{Im}\ln\det\Biggl[
            \sum_{\mu,\nu}^{N_{\rm basis}}
            C_{\nu n}^{\bm{k}_j*}\mathcal{S}_{\nu\mu}^{\bm{k}_j}
            e^{-i\Delta k\bm{b}_\alpha\cdot\bm{R}_\mu}
            C_{\mu m}^{\bm{k}_{j+1}}
        \Biggr]
        .
        \notag
    \end{align}
    This is the atomic-gauge expression implemented in practice to construct the cross-$\bm{k}$ overlaps entering the discretized Berry term.

        \section{Berry term matrix elements}\label{sec:berry}
To evaluate the second term in Eq.~(\ref{eq:psi-r-psi}), \ie the expectation value of the derivative with respect to the crystal momentum, we apply the product rule to Eq.~\eqref{eq:expansion}, which gives
    \begin{align}
        i\braket{\Psi_{n\bm{k}} \,|\,
        \frac{\partial\Psi_{n\bm{k}}}{\partial\bm{k}}}
        ={}&
        i\sum_{\mu,\nu}^{N_{\rm basis}}
        C_{\nu n}^{\bm{k}*}
        \mathcal{S}_{\nu\mu}^{\bm{k}}
        \frac{\partial C_{\mu n}^{\bm{k}}}{\partial\bm{k}}
        \label{eq:appendix-berry-product}\\
        &+i\sum_{\mu,\nu}^{N_{\rm basis}}
        C_{\nu n}^{\bm{k}*}C_{\mu n}^{\bm{k}}
        \braket{\chi_\nu^{\bm{k}} \,|\,
        \frac{\partial\chi_\mu^{\bm{k}}}{\partial\bm{k}}}.
        \notag
    \end{align}
    Using the Bloch sums in Eq.~\eqref{eq:basis-k-space}, the basis-function derivative is
    \begin{align}
        \braket{\chi_\nu^{\bm{k}} \,|\,
        \frac{\partial\chi_\mu^{\bm{k}}}{\partial\bm{k}}}
        ={}&
        \sum_{\bm{L}} i\bm{L}e^{i\bm{k}\cdot\bm{L}}
        \mathcal{S}_{\nu\mu}^{\bm{L}}
        =
        \frac{\partial\mathcal{S}_{\nu\mu}^{\bm{k}}}
        {\partial\bm{k}}.
        \label{eq:appendix-basis-derivative}
    \end{align}
    We therefore obtain
    \begin{align}
        \braket{\Psi_{n\bm{k}} \,|\,
        \frac{\partial\Psi_{n\bm{k}}}{\partial\bm{k}}}
        ={}&
        \sum_{\mu,\nu}^{N_{\rm basis}}
        C_{\nu n}^{\bm{k}*}
        \frac{\partial C_{\mu n}^{\bm{k}}}{\partial\bm{k}}
        \mathcal{S}_{\nu\mu}^{\bm{k}}
        \\
        &+\sum_{\mu,\nu}^{N_{\rm basis}}
        C_{\nu n}^{\bm{k}*}C_{\mu n}^{\bm{k}}
        \frac{\partial\mathcal{S}_{\nu\mu}^{\bm{k}}}
        {\partial\bm{k}} .\notag
    \end{align}
    The overlap-derivative term cancels the corresponding contribution from the dipole matrix elements derived in the supplementary material, Sec.~S3, and has therefore been omitted from Eq.~\eqref{eq:psi-der-k} of the main text.

    \bibliographystyle{aipnum4-2}
    \bibliography{bibliography}

\end{document}


\title{Supplementary Material for ``Scalable Berry-Phase Polarization Calculations with Numeric Atomic Orbitals''}
    \maketitle
    \onecolumngrid

    \renewcommand{\thesubsection}{S\arabic{subsection}}
    \renewcommand{\thefigure}{S\arabic{figure}}
    \renewcommand{\thetable}{S\arabic{table}}

    \subsection{Reciprocal vector derivative}\label{sec:k-derivative}

    The direct and reciprocal lattice vectors satisfy
    \begin{align}
        \bm{b}_\alpha\cdot\bm{a}_\beta=2\pi\delta_{\alpha\beta}. \label{eq:identity}
    \end{align}
    For a real-space vector $\bm{r}=\sum_\beta x^\beta\bm{a}_\beta$ within the primitive unit cell, where $x^\beta\in[0,1)$, Eq.~\eqref{eq:identity} gives $r^\alpha\equiv\bm{b}_\alpha\cdot\bm{r}=2\pi x^\alpha\in[0,2\pi)$.
    Denoting the Cartesian components of $\bm{a}_\beta$ and $\bm{b}_\beta$ by $a_{\beta i}$ and $b_{\beta i}$, respectively, the corresponding completeness relation is
    \begin{align}
        \sum_\beta \frac{a_{\beta i}b_{\beta j}}{2\pi}=\delta_{ij}
        \label{eq:reciprocal-completeness}
    \end{align}
    which follows by multiplying its left-hand side by \(a_{\gamma j}\), summing over \(j\), and using Eq.~\eqref{eq:identity}.
    Writing $\bm{k}=\sum_\alpha k_\alpha\bm{b}_\alpha$, an infinitesimal reciprocal-space displacement is $\mathrm{d}\bm{k}=\sum_\alpha \bm{b}_\alpha\,\mathrm{d}k_\alpha$.
    For any differentiable scalar function $f(\bm{k})$, its differential can therefore be written as
    \begin{align}
        \mathrm{d}f(\bm{k})
        &= \nabla_{\bm{k}}f(\bm{k})\cdot\mathrm{d}\bm{k} \\
        &= \sum_\alpha \underbrace{\left[\bm{b}_\alpha\cdot\nabla_{\bm{k}}f(\bm{k})\right]}_{\displaystyle\frac{\partial f(\bm{k})}{\partial k_\alpha}}\mathrm{d}k_\alpha.
        \label{eq:reciprocal-coordinate-derivative}
    \end{align}
    Multiplying the underbraced relation on the left by $\bm{a}_\beta/(2\pi)$ and summing over $\beta$, we obtain
    \begin{align}
        \sum_\beta \frac{\bm{a}_\beta}{2\pi}
        \frac{\partial f(\bm{k})}{\partial k_\beta}
        &= \sum_\beta \frac{\bm{a}_\beta}{2\pi}
        \left[\bm{b}_\beta\cdot\nabla_{\bm{k}}f(\bm{k})\right].
    \end{align}
    Expanding the right-hand side in Cartesian components and applying Eq.~\eqref{eq:reciprocal-completeness} yields
    \begin{align}
        \nabla_{\bm{k}}f(\bm{k}) = \sum_\alpha \frac{\bm{a}_\alpha}{2\pi}\frac{\partial f(\bm{k})}{\partial k_\alpha}.
        \label{eq:reciprocal-gradient}
    \end{align}

    \FloatBarrier

    \subsection{Matrix elements in reciprocal space}\label{sec:fourier}
    Given a local operator $\hat{A}$ that has the periodicity of the lattice, \ie
    \begin{align}
        \braket{\bm{r}' \,|\, \hat{A} \,|\, \bm{r}} =  \, A\left(\bm{r}\right)\delta\left( \bm{r}' -\bm{r}\right)
        \quad \text{and} \quad
        A\left(\bm{r} + \bm{L}\right) =  \,A\left(\bm{r}\right),
    \end{align}
    it is straightforward to express its matrix elements $A_{\nu\mu}^{\bm{k}}$ in the basis of Bloch-like wavefunctions $\chi_{\mu}^{\bm{k}}$ via a Fourier transform of the matrix elements $A_{\nu\mu}^{\bm{L}}$ of the same operator in the basis of real-space orbitals $\varphi_{\nu}^{\bm{L}}$:
    \begin{align}
        A_{\nu\mu}^{\bm{k}}
        =
        \braket{ \chi_{\nu}^{\bm{k}} \,|\, \hat{A} \,|\, \chi_{\mu}^{\bm{k}}  }
        =
        \sum_{\bm{L}}
        e^{i\bm{k}\cdot \bm{L}}
        \underbrace{
            \braket{ \varphi_{\nu}^{\bm{0}} \,|\, \hat{A} \,|\, \varphi_{\mu}^{\bm{L}} }
        }_{A_{\nu\mu}^{\bm{L}} }
    \end{align}
    In practice, it is advantageous to avoid evaluating the unbounded integral~$\braket{ \varphi_{\nu}^{\bm{0}} \,|\, \hat{A} \,|\, \varphi_{\mu}^{\bm{L}} }$ directly in a numerical implementation.
    For this reason, \aims evaluates this type of integral using a sum over lattice vectors:
    \begin{align}
    \braket{ \varphi_{\nu}^{\bm{0}} \,|\, \hat{A} \,|\, \varphi_{\mu}^{\bm{L}} }
    = \, \int \varphi_{\nu}^{\bm{0}*}({\bm r}) \, A(\bm{r}) \,
    \varphi_{\mu}^{\bm{L}}({\bm r}) \, d{\bm r}
    = \, \sum_{\bm L'} \int_{\rm u.c.}
    \varphi_{\nu}^{\bm{L'}*}({\bm r}) \, A(\bm{r}) \,  \varphi_{\mu}^{\bm{L}+{\bm L'}}({\bm r})
    \, d{\bm r},
    \end{align}
    which allows the actual integration to be restricted to the unit cell, provided that the operator $\hat{A}$ has the same periodicity as the lattice.
    The reciprocal-space representations of the overlap $\mathcal{S}^{\bm{k}}_{\nu\mu}$ and Hamiltonian $\mathcal{H}^{\bm{k}}_{\nu\mu}$ matrix elements are defined as
    \begin{align}
        \mathcal{S}^{\bm{k}}_{\nu\mu}
        = & \,
        \braket{ \chi_{\nu}^{\bm{k}} \,|\, \chi_{\mu}^{\bm{k}}  }
        =
        \sum_{\bm{L}}
        e^{i\bm{k}\cdot \bm{L}}
        \underbrace{
            \braket{ \varphi_{\nu}^{\bm{0}} \,|\, \varphi_{\mu}^{\bm{L}} }
        }_{\mathcal{S}^{\bm{L}}_{\nu\mu} }
        \\
        \mathcal{H}^{\bm{k}}_{\nu\mu}
        = & \,
        \braket{ \chi_{\nu}^{\bm{k}} \,|\, \hat{\mathcal{H}} \,|\, \chi_{\mu}^{\bm{k}}  }
        =
        \sum_{\bm{L}}
        e^{i\bm{k}\cdot \bm{L}}
        \underbrace{
            \braket{ \varphi_{\nu}^{\bm{0}} \,|\, \hat{\mathcal{H}} \,|\, \varphi_{\mu}^{\bm{L}} }
        }_{\mathcal{H}^{\bm{L}}_{\nu\mu} }
    \end{align}

    \FloatBarrier

    \subsection{Dipole matrix elements}\label{sec:dipole}
    The dipole matrix elements are defined through the expectation value of the position operator $\hat{\bm{r}}$, but we need to modify the procedure used in Sec.~\ref{sec:fourier}, because $\hat{\bm{r}}$ does not have the periodicity of the lattice.
    We start by defining the matrix elements $\bm{\mathcal{D}}^{\bm{L}}_{\nu\mu}$ of the position operator on the real-space localized orbitals (see Sec.~5.3 of Ref.~\cite{FHIaimsroadmap2025}) with respect to the center $\bm{R}_\mu+\bm{L}$ of the orbital $\varphi_{\mu}^{\bm{L}}$:
    \begin{align}
        \bm{\mathcal{D}}^{\bm{L}}_{\nu\mu} &= \braket{ \varphi_{\nu}^{\bm{0}} \,|\, \bm{r} - \bm{R}_\mu - \bm{L} \,|\, \varphi_{\mu}^{\bm{L}} }.
    \end{align}
    In the lattice gauge, substituting the Bloch sums of Eq.~(11) gives
    \begin{align}
        \braket{ \chi_{\nu}^{\bm{k}} \,|\, \bm{r} \,|\, \chi_{\mu}^{\bm{k}} } = \sum_{\bm{L}} e^{i\bm{k}\cdot\bm{L}} \left[ \bm{\mathcal{D}}^{\bm{L}}_{\nu\mu} + \left(\bm{R}_\mu+\bm{L}\right) \mathcal{S}^{\bm{L}}_{\nu\mu} \right] = \bm{\mathcal{D}}^{\bm{k}}_{\nu\mu} + \bm{R}_\mu\mathcal{S}^{\bm{k}}_{\nu\mu} -i\frac{\partial\mathcal{S}^{\bm{k}}_{\nu\mu}}{\partial\bm{k}}.
    \end{align}
    The overlap derivative originates from the lattice-vector term: in the Fourier sum, $\bm{L}e^{i\bm{k}\cdot\bm{L}}=-i\partial_{\bm{k}} e^{i\bm{k}\cdot\bm{L}}$.
    This term is not shown in Eq.~(15) in the main text for the sake of simplicity and clarity, since it cancels the overlap-derivative contribution to the Berry term derived in Appendix~B.

    Compared with a naive evaluation of $\braket{ \varphi_{\nu}^{\bm{0}} \,|\, \bm{r}  \,|\, \varphi_{\mu}^{\bm{L}} }$, this formulation uses a translationally invariant operator for the real-space integrals.
    As discussed in Sec.~\ref{sec:fourier}, this allows the use of the same numerical real-space integration routines employed to evaluate the Hamiltonian and overlap matrices:
    \begin{align}
        \bm{\mathcal{D}}^{\bm{L}}_{\nu\mu} = \,
        \int \varphi_{\nu}^{\bm{0}}({\bm r})^* \left( \bm{r} - \bm{R}_\mu - \bm{L} \right) \varphi_{\mu}^{\bm{L}}({\bm r}) \, d{\bm r}
        =  \,
        \sum_{\bm L'} \int_{\rm u.c.} \varphi_{\nu}^{\bm{L'}}({\bm r})^* \left( \bm{r} - \bm{R}_\mu - \bm{L}-\bm{L'} \right) \, \varphi_{\mu}^{\bm{L}+\bm{L'}}({\bm r}) \, d{\bm r}.
    \end{align}

    \FloatBarrier

    \subsection{Structural relaxations}\label{sec:geometries}

    The reference and polar phases used in the validation calculations were relaxed independently for each \gls{xc} functional.
    We optimized the atomic positions and cell parameters with the \texttt{BFGS} algorithm and the \texttt{FrechetCellFilter} implemented in \texttt{ase}~\cite{ase}.
    Symmetry constraints applied through \texttt{spglib}~\cite{spglib} preserved the assigned space group while allowing the internal coordinates and lattice parameters to relax.
    For \ce{PbTiO3}, \gls{soc} was not accounted for in the structural relaxation, but included perturbatively in the subsequent polarization calculations.
    For antiferromagnetic \ce{BiFeO3}, the Fe basis set was augmented with the $4p$ shell, and the two Fe sublattices were initialized with opposite magnetic moments.

    Tables~\ref{table:geometries-batio3-cubic-tetragonal}--\ref{table:geometries-pbtio} report the relaxed lattice and internal-site parameters for each material and phase, together with the literature structures used for comparison.
    In all Tables, a dash denotes a value not reported in the cited study.
    \tab{table:site-coordinate-conventions} defines the phase-specific fractional-coordinate conventions used for the internal-site parameters.
    These conventions are needed because the polar axes and symmetry-inequivalent oxygen sites differ among the three materials.

    In tetragonal $P4mm$ \ce{PbTiO3}, the polar axis is $[001]$ and Pb defines the origin.
    The two symmetry-equivalent equatorial oxygen atoms share one displacement parameter, while the apical oxygen uses a separate parameter.
    This convention follows the standard tetragonal description used, for example, in Ref.~\cite{btobecbis}.

    In rhombohedral $R3m$ \bto, the polar axis is $[111]$ and the three oxygen coordinates are related by permutation.
    The same fractional parameters are used for the theoretical and experimental structures.
    For \bto and \ce{PbTiO3}, we choose the symmetry-equivalent polar domain with a positive Ti displacement.

    In rhombohedral \ce{BiFeO3}, the Fe and oxygen coordinates describe both the polar displacement and the oxygen-octahedron rotation.
    The centrosymmetric and polar primitive cells use the same ten-atom coordinate convention, with the additional symmetry constraints of the centrosymmetric phase.
    Fractional coordinates are understood modulo one, and equivalent negative oxygen coordinates are used where convenient.

    \begin{table}[!htb]
    \centering
    \begin{tabular}{@{} l l l l l | l l | r r r @{}}
        \toprule
        system & crystal type & phase & Reference & \texttt{xc} & $a$ (\AA) & $c/a$ & \multicolumn{3}{c}{internal-site parameter} \\
        \midrule
        \bto & cubic & $Pm\bar{3}m$ & this work & \texttt{LDA} & 3.935 & 1.000 & \\
             &       &              &           & \texttt{PBEsol} & 3.968 & 1.000 & \\
             &       &              &           & \texttt{PBE} & 4.017 & 1.000 & \\
             &       &              &           & \texttt{HSE06} & 3.980 & 1.000 & \\
        \myline{2-10}
             & tetragonal & $P4mm$ & \multicolumn{4}{c|}{} & $dz_{\mathrm{Ti}}$ & $dz_{\mathrm{O}\perp}$ & $dz_{\mathrm{O}\parallel}$ \\
        \myline{2-10}
             & & & this work & \texttt{LDA} & 3.927 & 1.012 & 0.012 & 0.014 & 0.020 \\
             & & &           & \texttt{PBEsol} & 3.954 & 1.023 & 0.016 & 0.019 & 0.030 \\
             & & &           & \texttt{PBE} & 3.985 & 1.052 & 0.018 & 0.030 & 0.050 \\
             & & &           & \texttt{HSE06} & 3.952 & 1.047 & 0.019 & 0.026 & 0.047 \\
        \myline{4-10}
             & & & Ref.~\cite{choi2025first} & \texttt{LDA} & 3.93 & 1.01 & 0.012 & 0.014 & 0.021 \\
             & & & Ref.~\cite{zhang2017comparative} & \texttt{LDA} & 3.946 & 1.011 & 0.012 & - & - \\
             & & &           & \texttt{PBE} & 4.000 & 1.054 & 0.018 & - & - \\
             & & &           & \texttt{HSE03} & 3.959 & 1.039 & 0.019 & - & - \\
             & & & Ref.~\cite{choi2025first} & \texttt{PBEsol} & 3.96 & 1.02 & 0.016 & 0.019 & 0.030 \\
             & & & Ref.~\cite{choi2025first} & \texttt{HSE06} & 3.96 & 1.05 & 0.020 & 0.025 & 0.048 \\
             & & & Ref.~\cite{imura2014polarization} & Exp. & 3.990 & 1.011 & - & - & - \\
        \bottomrule
    \end{tabular}
    \caption{Relaxed lattice and internal-site parameters for cubic and tetragonal \bto, together with literature structures.
    Tables~S1--S4 use the phase-specific fractional-coordinate conventions defined in \tab{table:site-coordinate-conventions} and list a dash when a cited study does not report a value.
    Blank internal-site entries for cubic phases are fixed to zero by symmetry.}
    \label{table:geometries-batio3-cubic-tetragonal}
    \alttext{%
        Structural parameters of cubic and tetragonal \bto.
    }
\end{table}

    \begin{table}[!htb]
    \centering
    \begin{tabular}{@{} l l l l l | l l | r r r r @{}}
        \toprule
        system & crystal type & phase & Reference & \texttt{xc} & $a$ (\AA) & $\alpha$ (${}^\circ$) & \multicolumn{4}{c}{internal-site parameter} \\
        \midrule
        \bto & rhombohedral & $R3m$ & \multicolumn{4}{c|}{} & $\Delta r_{\mathrm{Ti}}$ & $\Delta r_{\mathrm{O1}}$ & $\Delta r_{\mathrm{O2}}$ & \\
        \myline{2-11}
             & & & this work & \texttt{LDA} & 3.945 & 89.93 & 0.009 & 0.009 & 0.013 & \\
             & & &           & \texttt{PBEsol} & 3.988 & 89.86 & 0.011 & 0.012 & 0.018 & \\
             & & &           & \texttt{PBE} & 4.057 & 89.70 & 0.014 & 0.016 & 0.027 & \\
             & & &           & \texttt{HSE06} & 4.016 & 89.74 & 0.014 & 0.014 & 0.025 & \\
        \myline{4-11}
             & & & Ref.~\cite{seo2013polytypes} & \texttt{LDA} & 3.962 & 89.961 & - & - & - & \\
             & & &           & \texttt{PBE} & 4.065 & 89.755 & - & - & - & \\
             & & & Ref.~\cite{goh2016structural} & \texttt{PBE} & 4.001 & 89.87 & 0.011 & 0.011 & 0.017 & \\
             & & & Ref.~\cite{hewat1974structure}\textsuperscript{\(\dagger\)} & Exp. & 4.001 & 89.868(4) & 0.013(3) & 0.011(4) & 0.018(4) & \\
        \bottomrule
    \end{tabular}
    \caption{Relaxed lattice and internal-site parameters for rhombohedral \bto, together with literature structures.
    For shared conventions and notation, see the caption of Table~\ref{table:geometries-batio3-cubic-tetragonal}.
    Ref.~\cite{wieder1955electrical}, which is shown in Table~I, does not report lattice parameters, so the low-temperature neutron-diffraction structure from Ref.~\cite{hewat1974structure} is listed instead here.
    References marked with \textsuperscript{\(\dagger\)} do not report a polarization and are not included in Table~I.}
    \label{table:geometries-batio3-rhombohedral}
    \alttext{%
        Structural parameters of rhombohedral \bto.
    }
\end{table}

    \begin{table}[!htb]
    \centering
    \begin{tabular}{@{} l l l l l | r r | r r r r @{}}
        \toprule
        system & crystal type & phase & Reference & \texttt{xc} & $a$ (\AA) & $\alpha$ (${}^\circ$) & \multicolumn{4}{c}{internal-site parameter} \\
        \midrule
        \ce{BiFeO3} & & \multicolumn{5}{c|}{} & $x_{\mathrm{Fe}}$ & $x_{\mathrm{O}}$ & $y_{\mathrm{O}}$ & $z_{\mathrm{O}}$ \\
        \myline{2-11}
        & rhombohedral & $R\bar{3}c$ & this work & \texttt{LDA} & 5.351 & 61.87 & 0.250 & 0 & 0.083 & 0.083 \\
        &              &              &           & \texttt{PBEsol} & 5.411 & 61.70 & 0.250 & 0 & 0.083 & 0.083 \\
        &              &              &           & \texttt{PBE} & 5.507 & 61.38 & 0.250 & 0 & 0.085 & 0.085 \\
        &              &              &           & \texttt{HSE06} & 5.476 & 61.18 & 0.250 & 0 & 0.083 & 0.083 \\
        \myline{2-11}
        & rhombohedral & $R3c$ & this work & \texttt{LDA} & 5.460 & 60.27 & 0.230 & 0.057 & 0.041 & 0.102 \\
        &              &       &           & \texttt{PBEsol} & 5.544 & 59.90 & 0.228 & 0.059 & 0.039 & 0.106 \\
        &              &       &           & \texttt{PBE} & 5.685 & 59.20 & 0.223 & 0.064 & 0.033 & 0.113 \\
        &              &       &           & \texttt{HSE06} & 5.649 & 59.20 & 0.221 & 0.067 & 0.026 & 0.105 \\
        \myline{4-11}
        &              &       & Ref.~\cite{choi2025first} & \texttt{LDA} & 5.466 & 60.171 & - & - & - & - \\
        &              &       &                           & \texttt{PBEsol} & 5.556 & 59.811 & - & - & - & - \\
        &              &       &                           & \texttt{HSE06} & 5.647 & 59.102 & - & - & - & - \\
        &              &       & Ref.~\cite{zhang2017comparative} & \texttt{LDA} & 5.442 & 60.444 & - & - & - & - \\
        &              &       &                           & \texttt{PBE} & 5.685 & 59.225 & - & - & - & - \\
        &              &       &                           & \texttt{HSE03} & 5.651 & 59.119 & - & - & - & - \\
        &              &       & Ref.~\cite{bfo2005} & \texttt{LDA} & 5.460 & 60.36 & 0.231 & 0.057 & 0.042 & 0.102 \\
        &              &       & Ref.~\cite{li2004dramatically} & Exp. & 5.630 & 59.31 & - & - & - & - \\
        \bottomrule
    \end{tabular}
    \caption{Relaxed lattice and internal-site parameters for \ce{BiFeO3}, together with literature structures.
    For shared conventions and notation, see the caption of Table~\ref{table:geometries-batio3-cubic-tetragonal}.
    The experimental pseudocubic cell from Ref.~\cite{li2004dramatically} is converted to the 10-atom rhombohedral primitive-cell convention used here.}
    \label{table:geometries-bifeo3}
    \alttext{%
        Structural parameters of \ce{BiFeO3}.
    }
\end{table}

    \begin{table}[!htb]
    \centering
    \begin{tabular}{@{} l l l l l | r r | r r r @{}}
        \toprule
        system & crystal type & phase & Reference & \texttt{xc} & $a$ (\AA) & $c/a$ & \multicolumn{3}{c}{internal-site parameter} \\
        \midrule
        \ce{PbTiO3} & & \multicolumn{5}{c|}{} & $dz_{\mathrm{Ti}}$ & $dz_{\mathrm{O}\perp}$ & $dz_{\mathrm{O}\parallel}$ \\
        \myline{2-10}
        & cubic & $Pm\bar{3}m$ & this work & \texttt{LDA} & 3.885 & 1.000 & \\
        &       &              &           & \texttt{PBEsol} & 3.914 & 1.000 & \\
        &       &              &           & \texttt{PBE} & 3.965 & 1.000 & \\
        &       &              &           & \texttt{HSE06} & 3.926 & 1.000 & \\
        \myline{2-10}
        & tetragonal & $P4mm$ & this work & \texttt{LDA} & 3.857 & 1.048 & 0.033 & 0.104 & 0.092 \\
        &            &         &           & \texttt{PBEsol} & 3.873 & 1.073 & 0.035 & 0.115 & 0.109 \\
        &            &         &           & \texttt{PBE} & 3.840 & 1.239 & 0.057 & 0.168 & 0.192 \\
        &            &         &           & \texttt{HSE06} & 3.813 & 1.216 & 0.054 & 0.163 & 0.183 \\
        \myline{4-10}
        &            &         & Ref.~\cite{choi2025first} & \texttt{LDA} & 3.86 & 1.05 & 0.034 & 0.105 & 0.093 \\
        &            &         &                           & \texttt{PBEsol} & 3.86 & 1.10 & 0.040 & 0.128 & 0.126 \\
        &            &         &                           & \texttt{HSE06} & 3.81 & 1.22 & 0.056 & 0.166 & 0.186 \\
        &            &         & Ref.~\cite{zhang2017comparative} & \texttt{LDA} & 3.865 & 1.045 & 0.034 & - & - \\
        &            &         &                           & \texttt{PBE} & 3.844 & 1.240 & 0.058 & - & - \\
        &            &         &                           & \texttt{HSE03} & 3.832 & 1.158 & 0.047 & - & - \\
        &            &         & Ref.~\cite{saghi1999first} & \texttt{PBE} & 3.893 & 1.073 & - & - & - \\
        &            &         & Ref.~\cite{pto2004} & Exp. & 3.904 & 1.063 & - & - & - \\
        \bottomrule
    \end{tabular}
    \caption{Relaxed lattice and internal-site parameters for \ce{PbTiO3}, together with literature structures.
    For shared conventions and notation, see the caption of Table~\ref{table:geometries-batio3-cubic-tetragonal}.}
    \label{table:geometries-pbtio}
    \alttext{%
        Structural parameters of \ce{PbTiO3}.
    }
\end{table}

    \begin{table}[!htb]
    \centering
    \begin{tabular}{@{} l l l l l l @{}}
        \toprule
        system         & atom    & $x$                          & $y$                          & $z$                          \\
        \midrule
        \bto    & Ba      & $0$                          & $0$                          & $0$                          \\
        (tetragonal)   & Ti      & $0.5$                        & $0.5$                        & $0.5+dz_{\mathrm{Ti}}$       \\
                       & \ce{O1} & $0.5$                        & $0$                          & $0.5-dz_{\mathrm{O}\perp}$   \\
                       & \ce{O2} & $0$                          & $0.5$                        & $0.5-dz_{\mathrm{O}\perp}$   \\
                       & \ce{O3} & $0.5$                        & $0.5$                        & $\phantom{0.5}-dz_{\mathrm{O}\parallel}$  \\
        \midrule
        \bto    & Ba      & $0$                          & $0$                          & $0$                          \\
        (rhombohedral) & Ti      & $0.5+\Delta r_{\mathrm{Ti}}$ & $0.5+\Delta r_{\mathrm{Ti}}$ & $0.5+\Delta r_{\mathrm{Ti}}$ \\
                       & \ce{O1} & $0.5-\Delta r_{\mathrm{O1}}$ & $0.5-\Delta r_{\mathrm{O1}}$ & $\phantom{0.5}-\Delta r_{\mathrm{O2}}$    \\
                       & \ce{O2} & $0.5-\Delta r_{\mathrm{O1}}$ & $\phantom{0.5}-\Delta r_{\mathrm{O2}}$    & $0.5-\Delta r_{\mathrm{O1}}$ \\
                       & \ce{O3} & $\phantom{0.5}-\Delta r_{\mathrm{O2}}$    & $0.5-\Delta r_{\mathrm{O1}}$ & $0.5-\Delta r_{\mathrm{O1}}$ \\
        \midrule
        \ce{BiFeO3}    & Bi$_1$  & $0$                          & $0$                          & $0$                          \\
                       & Bi$_2$  & $0.5$                        & $0.5$                        & $0.5$                        \\
                       & Fe$_1$  & $\phantom{0.5}+x_{\mathrm{Fe}}$            & $\phantom{0.5}+x_{\mathrm{Fe}}$            & $\phantom{0.5}+x_{\mathrm{Fe}}$            \\
                       & Fe$_2$  & $0.5+x_{\mathrm{Fe}}$        & $0.5+x_{\mathrm{Fe}}$        & $0.5+x_{\mathrm{Fe}}$        \\
                       & O$_1$   & $0.5-x_{\mathrm{O}}$             & $\phantom{0.5}+y_{\mathrm{O}}$             & $\phantom{0.5}-z_{\mathrm{O}}$             \\
                       & O$_2$   & $\phantom{0.5}-z_{\mathrm{O}}$             & $0.5-x_{\mathrm{O}}$             & $\phantom{0.5}+y_{\mathrm{O}}$             \\
                       & O$_3$   & $\phantom{0.5}+y_{\mathrm{O}}$             & $\phantom{0.5}-z_{\mathrm{O}}$             & $0.5-x_{\mathrm{O}}$             \\
                       & O$_4$   & $\phantom{0.5}-x_{\mathrm{O}}$         & $0.5-z_{\mathrm{O}}$         & $0.5+y_{\mathrm{O}}$         \\
                       & O$_5$   & $0.5-z_{\mathrm{O}}$         & $0.5+y_{\mathrm{O}}$         & $\phantom{0.5}-x_{\mathrm{O}}$         \\
                       & O$_6$   & $0.5+y_{\mathrm{O}}$         & $\phantom{0.5}-x_{\mathrm{O}}$         & $0.5-z_{\mathrm{O}}$         \\
        \midrule
        \ce{PbTiO3}    & Pb      & $0$                          & $0$                          & $0$                          \\
                       & Ti      & $0.5$                        & $0.5$                        & $0.5+dz_{\mathrm{Ti}}$       \\
                       & \ce{O1} & $0.5$                        & $0$                          & $0.5+dz_{\mathrm{O}\perp}$   \\
                       & \ce{O2} & $0$                          & $0.5$                        & $0.5+dz_{\mathrm{O}\perp}$   \\
                       & \ce{O3} & $0.5$                & $0.5$                & $\phantom{0.5}+dz_{\mathrm{O}\parallel}$   \\
        \bottomrule
    \end{tabular}
    \caption{
    Fractional atomic coordinates of the distorted validation structures are defined in terms of the internal-site parameters reported in Tables~\ref{table:geometries-batio3-cubic-tetragonal}--\ref{table:geometries-pbtio}.
    All ten atoms in the rhombohedral \ce{BiFeO3} primitive cell are shown explicitly, and all coordinates are understood modulo one.}
    \label{table:site-coordinate-conventions}
    \alttext{%
        Fractional atomic coordinates.
    }
\end{table}

    \FloatBarrier
    \clearpage

    \subsection{Timing components of the \lapack calculation}\label{sec:lapack-components}

\begin{figure}[!htb]
    \centering
    \includegraphics[width=0.72\linewidth]{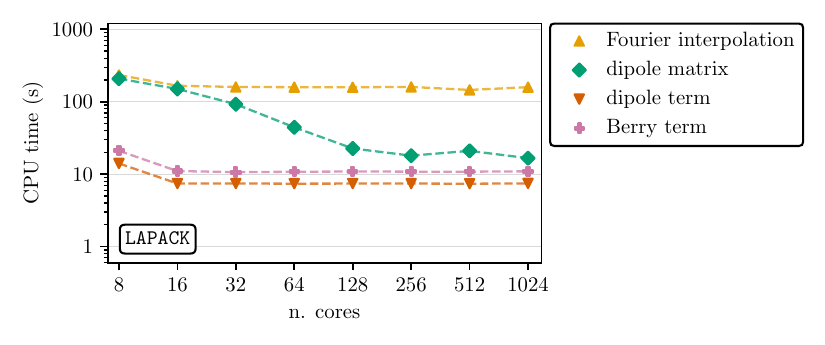}
    \caption{Timing components of the \lapack polarization calculation.}
    \label{fig:lapack-component-timings}
    \alttext{%
        Component CPU times versus number of cores.
    }
\end{figure}

    Figure~\ref{fig:lapack-component-timings} resolves the internal timing components of the \lapack calculation for the \ce{MgO} benchmark of Fig.~5.
    The \scf and \nscf $\bm{k}$-grids are $2\times2\times2$ and $2\times2\times16$, respectively, with the polarization evaluated along the \nth{3} reciprocal-lattice direction.
    The reported components are internal wall-clock timers, which sum up to the net CPU polarization time shown in Fig.~5.
    Here, we indeed observe that the scalability of the Fourier interpolation step reaches a plateau at 16 cores, \ie when the number of cores equals the number of $\bm{k}$-points in each string.
    The same logic holds for the Berry- and dipole-term contributions.
    For larger core counts, some cores are idle and no longer contribute to the evaluation of these terms, and only the construction of the real-space dipole matrix continues to become less expensive as the number of cores grows.
    Because the dipole-matrix construction makes up for a substantial fraction of the polarization cost, its continued reduction delays the plateau of the total \lapack time.

    \FloatBarrier
    \clearpage

    \subsection{Timing profile of polarization costs}\label{sec:breakdown}

\begin{figure*}[!ht]
    \centering
    \includegraphics[width=\linewidth]{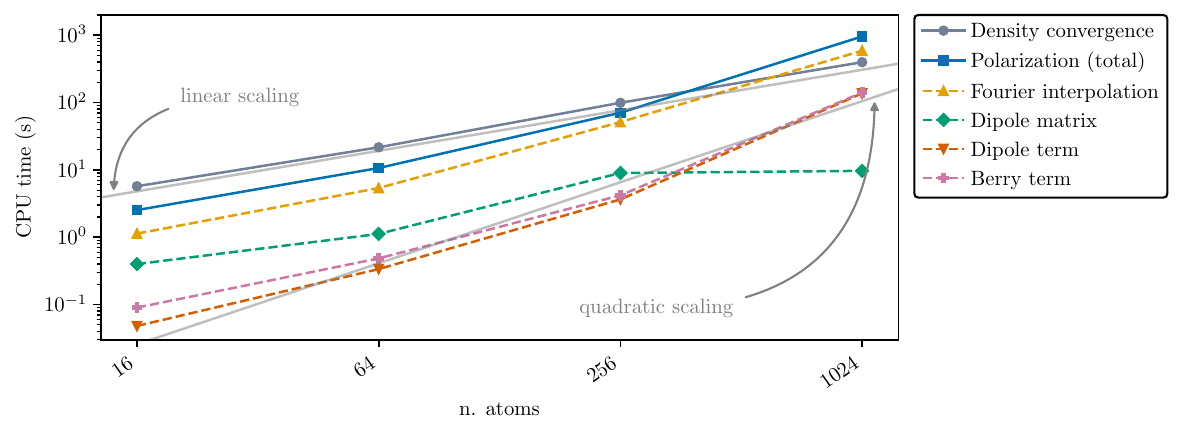}
    \caption{%
    Net CPU time for polarization and its contributions for \ce{MgO} supercells using 512 cores, together with the time required to converge the charge density.
    The gray lines indicate linear and quadratic reference dependence on the number of atoms and are not fits to the data.
    }
    \label{fig:scalability-components}
    \alttext{%
        CPU times versus number of atoms.
    }
\end{figure*}

    \fig{fig:scalability-components} presents a timing profile of the reported \scalapack polarization time as a function of system size.
    The solid blue curve is the same net \scalapack polarization time shown in Figure~6.
    The Fourier-interpolation, dipole-matrix, dipole-term, and Berry-term curves report internal \aims timers for the principal stages of the calculation.
    The Fourier interpolation is the most time-consuming operation throughout and becomes increasingly dominant for larger systems, since it includes the computationally dominant eigensolver, which formally scales cubically with system size.
    The dipole and Berry terms increase in computational cost with system size but remain substantially less expensive.
    For smaller systems, computing the polarization is considerably cheaper than converging the charge density and therefore represents only a modest additional cost.
    As the system size increases, the polarization cost grows relative to the charge-density-convergence cost, and the two become comparable for the larger systems considered here.

    \begin{table}[!htb]
    \centering
    \begin{tabular}{@{}r c c c@{}}
        \toprule
        \text{atoms} & \text{supercell} & \scf $\bm{k}$-grid & \nscf $\bm{k}$-grid \\
        \midrule
        16   & $2\times2\times2$ & $8\times8\times8$ & $32\times8\times8$ \\
        64   & $4\times4\times2$ & $4\times4\times8$ & $16\times4\times8$ \\
        256  & $8\times4\times4$ & $2\times4\times4$ & $8\times4\times4$ \\
        1024 & $8\times8\times8$ & $2\times2\times2$ & $8\times2\times2$ \\
        \bottomrule
    \end{tabular}
    \caption{The \scf and \nscf $\bm{k}$-grids used for the \scalapack timing data in Fig.~\ref{fig:scalability-components}.}
    \label{tab:scalapack-atom-scaling-grids}
    \alttext{%
        Supercells and $\bm{k}$-grids.
    }
\end{table}

    \FloatBarrier
    \clearpage

    \subsection{Component scalability analysis}\label{sec:component-scalability}

\begin{figure*}[!ht]
    \centering
    \includegraphics[width=\linewidth]{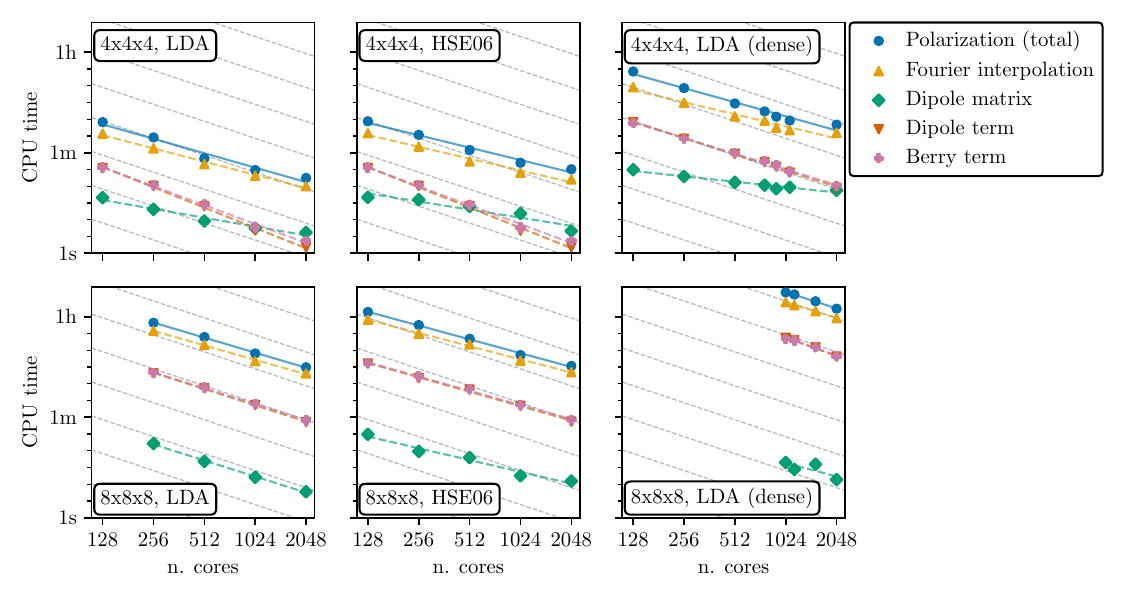}
    \caption{Component timings as a function of core count for the $4\times4\times4$ supercell (top row) and the $8\times8\times8$ supercell (bottom row) of \bto.
    The panels show the total polarization time, Fourier interpolation, dipole-matrix construction, dipole term, and Berry term; gray lines indicate ideal scalability.}
    \label{fig:component-scaling-ideal}
    \alttext{%
        Component CPU times versus number of cores.
    }
\end{figure*}

    We analyze the parallel scalability of the principal contributions to the \scalapack polarization calculation for the \bto $4\times4\times4$ and $8\times8\times8$ supercells considered in the main text.
    The \scf and \nscf $\bm{k}$-grids used for these data sets are listed in Table~\ref{tab:component-scalability-grids}.
    Figure~\ref{fig:component-scaling-ideal} provides the component-resolved breakdown of the four data sets shown in Fig.~5, together with two additional \texttt{LDA} data sets for the same supercells calculated using denser $\bm{k}$-grids.

    \begin{table}[!thb]
    \centering
    \begin{tabular}{@{}r c c c c@{}}
        \toprule
        \text{atoms} & \text{supercell} & xc & \scf $\bm{k}$-grid & \nscf $\bm{k}$-grid \\
        \midrule
        320  & $4\times4\times4$ & \texttt{LDA}             & $2\times2\times2$ & $2\times2\times4$ \\
             &                   & \texttt{HSE06}           & $2\times2\times2$ & $2\times2\times4$ \\
             &                   & \texttt{LDA} (dense)     & $4\times4\times4$ & $4\times4\times6$ \\
        \arrayrulecolor{gray}
        \midrule[\lightrulewidth]
        \arrayrulecolor{black}
        2560 & $8\times8\times8$ & \texttt{LDA}             & $1\times1\times1$ & $1\times1\times2$ \\
             &                   & \texttt{HSE06}           & $1\times1\times1$ & $1\times1\times2$ \\
             &                   & \texttt{LDA} (dense)     & $2\times2\times2$ & $2\times2\times3$ \\
        \bottomrule
    \end{tabular}
    \caption{The \scf and \nscf $\bm{k}$-grids used for the data in Fig.~\ref{fig:component-scaling-ideal}.
    The \nscf $\bm{k}$-grid is used to calculate the polarization along the \nth{3} lattice direction.}
    \label{tab:component-scalability-grids}
    \alttext{%
        Supercells and $\bm{k}$-grids.
    }
\end{table}

    \FloatBarrier

    \FloatBarrier
    \clearpage

    \subsection{Parallel scalability for liquid-water polarization}\label{sec:water-scalability}

\begin{figure*}[!hb]
    \centering
    \includegraphics[width=\linewidth]{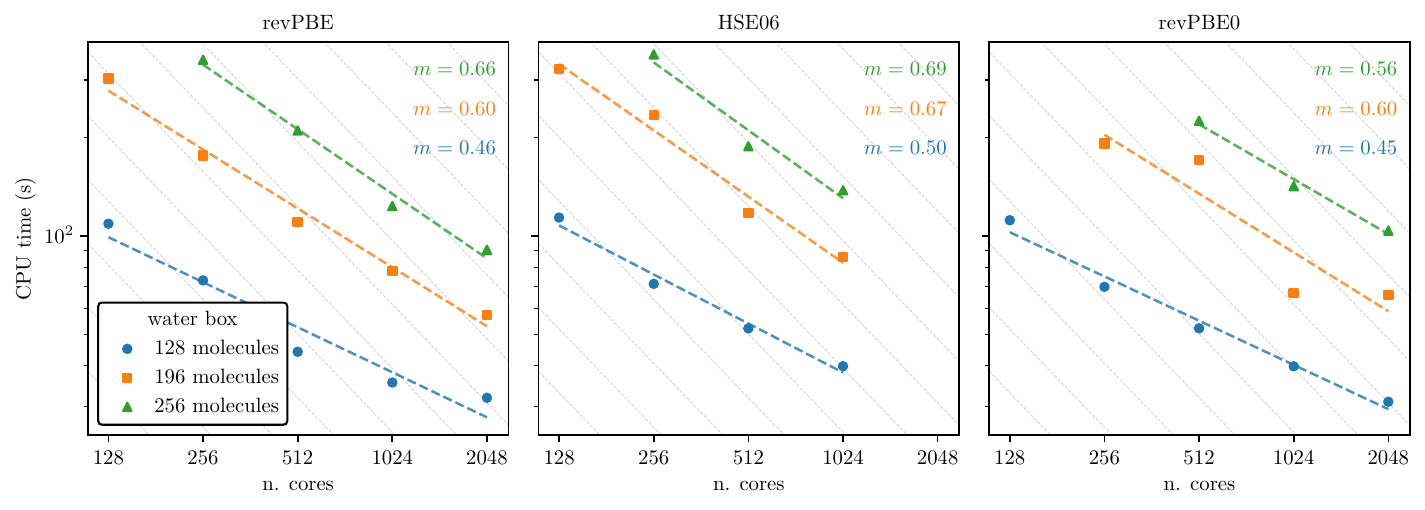}
    \caption{%
    Net polarization CPU time as a function of core count for liquid-water boxes containing 128, 196, and 256 molecules, calculated with the \texttt{revPBE} (left), \texttt{HSE06} (center), and \texttt{revPBE0} (right) \gls{xc} functionals.
    Colored dashed lines are power-law fits, and gray dashed lines indicate ideal scalability.
    }
    \label{fig:water-scalability}
    \alttext{%
        CPU time versus number of cores for liquid water.
    }
\end{figure*}

    \fig{fig:water-scalability} compares the parallel scalability of liquid-water polarization calculations for three water-box sizes for each of the \texttt{revPBE}, \texttt{HSE06}, and \texttt{revPBE0} \gls{xc} functionals.
    The polarization time decreases throughout the sampled core-count range for every functional and box size.
    The fitted exponents range from $m=0.45$ to $m=0.69$, indicating subideal scalability due to the small size of the system.
    The exponents are similar, at least for the two largest boxes, for the three functionals, indicating that the scalability is independent of the chosen functional for large enough systems.

    \FloatBarrier

    \subsection{\texorpdfstring{Parallel scalability with denser $\bm{k}$-grids}{Parallel scalability with denser k-grids}}\label{sec:dense-k-grid-scalability}
\begin{figure}[!htb]
    \centering
    \includegraphics[width=0.85\linewidth]{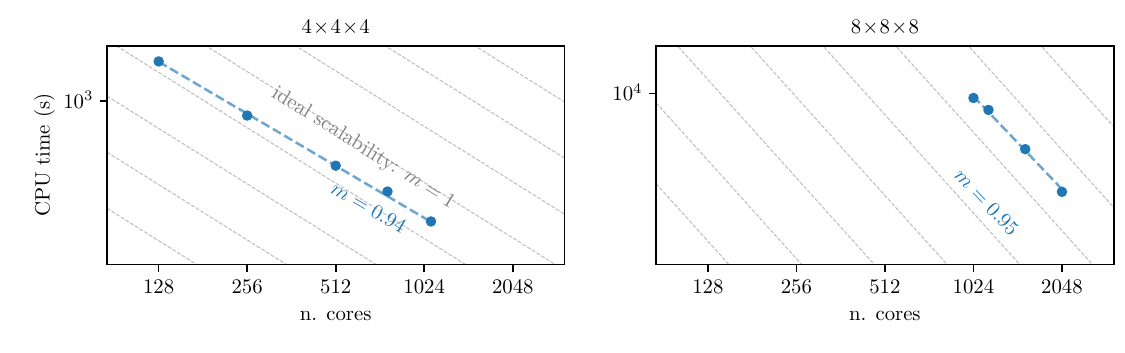}
    \caption{Net polarization time as a function of the number of cores for the $4\times4\times4$ and $8\times8\times8$ \bto supercells calculated with \texttt{LDA} and denser \scf and \nscf $\bm{k}$-grids.
    The gray dashed lines indicate ideal scalability ($m=1$), while the blue dashed lines are power-law fits.}
    \label{fig:dense-k-grid-scalability}
    \alttext{%
        CPU time versus number of cores.
    }
\end{figure}

The \texttt{LDA} calculations for the $4\times4\times4$ supercell used $4\times4\times4$ and $4\times4\times6$ \scf and \nscf grids, respectively, while those for the $8\times8\times8$ supercell used $2\times2\times2$ and $2\times2\times3$ grids.
The net polarization timings and power-law fits are shown in \fig{fig:dense-k-grid-scalability}; the fitted exponents are $m=0.94(2)$ and $m=0.95(4)$ for the smaller and larger supercells, respectively.

    \FloatBarrier
    \clearpage

    \bibliographystyle{aipnum4-2}
    \bibliography{bibliography}